\documentclass[%
preprint,
superscriptaddress,
amsmath,amssymb,
aps,
longbibliography,
]{revtex4-1}

\usepackage{graphicx}% Include figure files
\usepackage{dcolumn}% Align table columns on decimal point
\usepackage{bm}% bold math
\usepackage{xcolor}
\usepackage[normalem]{ulem}

\newcommand{\new}[1]{\textcolor{black}{{}#1}}

\usepackage[pagebackref=false]{hyperref}

\begin{document}
	
	\author{L.~Banszerus}
	\email{luca.banszerus@rwth-aachen.de}
	\author{S.~M\"oller}
	\affiliation{JARA-FIT and 2nd Institute of Physics, RWTH Aachen University, 52074 Aachen, Germany,~EU}%
	\affiliation{Peter Gr\"unberg Institute  (PGI-9), Forschungszentrum J\"ulich, 52425 J\"ulich,~Germany,~EU}
	\author{C.~Steiner}
	\author{E.~Icking}
	\affiliation{JARA-FIT and 2nd Institute of Physics, RWTH Aachen University, 52074 Aachen, Germany,~EU}%
	\affiliation{Peter Gr\"unberg Institute  (PGI-9), Forschungszentrum J\"ulich, 52425 J\"ulich,~Germany,~EU}
	
	\author{S.~Trellenkamp}
	\author{F.~Lentz}
	\affiliation{Helmholtz Nano Facility, Forschungszentrum J\"ulich, 52425 J\"ulich,~Germany,~EU}
	
	\author{K.~Watanabe}
	\affiliation{Research Center for Functional Materials, 
		National Institute for Materials Science, 1-1 Namiki, Tsukuba 305-0044, Japan}
	\author{T.~Taniguchi}
	\affiliation{International Center for Materials Nanoarchitectonics, 
		National Institute for Materials Science,  1-1 Namiki, Tsukuba 305-0044, Japan}%
	
	\author{C.~Volk}
	\author{C.~Stampfer}
	\affiliation{JARA-FIT and 2nd Institute of Physics, RWTH Aachen University, 52074 Aachen, Germany,~EU}%
	\affiliation{Peter Gr\"unberg Institute  (PGI-9), Forschungszentrum J\"ulich, 52425 J\"ulich,~Germany,~EU}%

	\title{Spin-valley coupling in single-electron bilayer graphene quantum dots}
	
	\date{\today}% It is always \today, today,
	%  but any date may be explicitly specied

	\keywords{quantum dot, bilayer graphene, Kane-Mele, spin-orbit, double quantum dot}
	
	\begin{abstract}

		\textbf{Understanding how the electron spin is coupled to orbital degrees of freedom, such as a valley degree of freedom  in solid-state systems 
			is central to applications in spin-based electronics and quantum computation. 
			Recent developments in the preparation of electrostatically-confined quantum dots in gapped bilayer graphene (BLG) enable to study the low-energy single-electron spectra in BLG quantum dots, which is crucial for potential spin and spin-valley qubit operations. Here, we present the observation of the spin-valley coupling in bilayer graphene \new{quantum dots} in the single-electron regime. 
			By making use of highly-tunable double quantum dot \new{devices} we achieve an energy resolution allowing us to resolve the lifting of the fourfold spin and valley degeneracy by a Kane-Mele type spin-orbit coupling of \new{$\approx 60~\mu$eV}.
			Furthermore, we find an upper limit of a potentially disorder-induced mixing of the $K$ and $K'$ states below $20~\mu$eV.}
		
	\end{abstract}
	
	\maketitle
	
\begin{figure}[!thb]
		\includegraphics[draft=false,keepaspectratio=true,clip,width=0.5
		\linewidth]{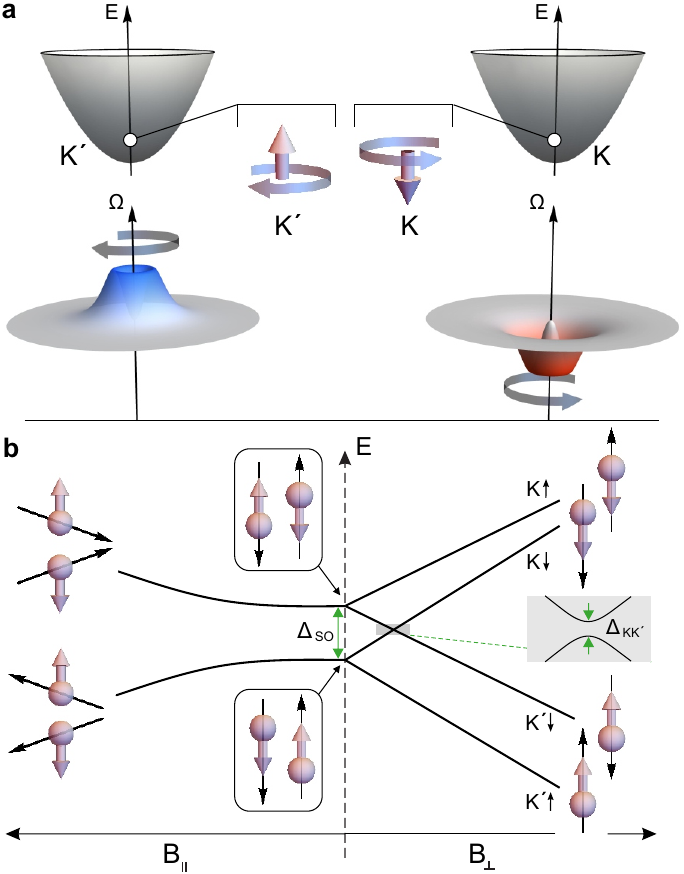}
		\caption[Fig01]{
			\textbf{a} Low energy band schematic of BLG at the $K$ and $K'$ points. 
			BLG exhibits a non-trivial Berry curvature $\Omega$ that leads to an effective out-of-plane magnetic moment with opposite sign at $K$ and $K'$. 
			\textbf{b} 
			Energy dispersion of single-particle states in BLG QDs as a function of in-plane ($B_\parallel$, left) and out-of-plane ($B_\perp$, right) applied magnetic fields with respect to the BLG plane.
			The SO gap, $\Delta_\mathrm{SO}$, lifts the fourfold degeneracy and polarizes the spins out-of-plane for zero magnetic field and a potential $K$-$K'$ state mixing (described by $\Delta_\mathrm{KK'}$) leads to an anticrossing of the \new{$K\!\!\downarrow$ and $K'\!\!\downarrow$ state}.}
		\label{f1}
	\end{figure}

	The valley pseudospin is an inherent property of two-dimensional honeycomb crystals and - together with the electron spin - makes graphene and bilayer graphene (BLG) interesting for applications in spin- and valley-based electronics and quantum computation~\cite{Trauzettel2007Mar,Wu2013Sep}. 
	This %valley 
	pseudospin arises form the orbital degree of freedom of the %degenerate and
	independent energy valleys located  at the  inequivalent  vertices ($K$ and $K'$) of the hexagonal Brillouin zone~\cite{McCann2013Apr}.
	In analogy to the real spin, the valley pseudospin exhibits also a valley Zeeman effect~\cite{Knothe2018Oct,Eich2018Aug,Banszerus2020Oct}, where the valley Zeeman splitting -- varying linearly with (out-of-plane) magnetic field -- is a result of the orbital magnetic moments originating from the non-vanishing Berry curvature, $\Omega$, at the K-points of gapped BLG (see Fig.~\ref{f1}a). %, which have opposite sign for the two valleys. \\
	Since these magnetic moments, which have opposite signs for the two valleys, crucially depend on the wave function, the valley $g$-factor in BLG quantum dots can be tuned by electric fields~\cite{Lee2020Mar,Garreis2020Nov}, offering promising and interesting possibilities for manipulation. 
	However, to fully exploit the potential to manipulate and control both the valley and spin degrees of freedom in BLG quantum dots (QDs), a detailed understanding of their interaction is essential.
	This is as relevant for a better understanding of spin decoherence processes as it is for exploring ways to electrically manipulate the spin degree of freedom via spin-orbit interaction and implementing innovative spin-valley qubits~\cite{Wu2013Sep}.
	Indeed, a detailed understanding of the low-energy spectrum of single particle states within the first electronic orbital (see Fig.~1b) is crucial for finding suitable working points and manipulation mechanisms for possible qubit operation. 
	
	Although the single-particle spectrum in BLG QDs has been intensively studied in recent years~\cite{Eich2018Jul,Kurzmann2019Jul,Knothe2020Jun}, the low-energy spin-valley coupling in BLG QDs has remained experimentally unexplored. This is certainly partly due to the high energy resolution required, as theoretical studies predict an intrinsic spin-orbit (SO) coupling in graphene and BLG of around $\Delta_{\mathrm{SO}} \sim 24\,\mu$eV \cite{Konschuh2012Mar, Kane2005Nov, Min2006Oct, Yao2007Jan, Zollner2019Mar}
	and only recently, experiments have -- partly indirectly -- reported values in the range between $40$ and~$80\,\mu$eV~\cite{Banszerus2020May,Sichau2019Feb}. 
	Moreover, our current knowledge with respect to a possible mixing of $K$ and $K'$ states is very limited. The latter is expressed by $\Delta_\mathrm{KK'}$ and could allow to access helical states~\cite{Klinovaja2012Dec}.

	In this letter, we report on measurements of the excited state spectrum of  single-electron double quantum \new{dots (DQDs)} in BLG providing information on $\Delta_\mathrm{SO}$ as well as on $\Delta_\mathrm{KK'}$. By tuning \new{a DQD} to a regime of low interdot tunnel coupling, we are able to resolve the interdot transitions with remarkably high energy resolution allowing to reconstruct the underlying single particle spectrum of both quantum dots. We find that the spin and valley degeneracy of the single particle spectrum is lifted by a Kane-Mele type SO gap~\cite{Kane2005Nov} of \new{$\Delta_\mathrm{SO} \approx 60$~$\mu$eV}, which separates the two Kramer's pairs -- $(K'\!\!\uparrow, K\!\!\downarrow)$ and $(K'\!\!\downarrow, K\!\!\uparrow)$ -- similar (but smaller in magnitude) to what has been observed in carbon nanotube QDs~\cite{Kuemmeth2008Mar,Jespersen2011Apr}. 
	The disorder-induced mixing of $K$ and $K'$ \new{states} is found to be at least smaller than $\Delta_\mathrm{KK'}<20$~$\mu$eV, \new{where the upper bound is resulting from the energy resolution of our measurements}. 
	Fig.~\ref{f1}b depicts the first four BLG QD states composing the first electronic orbital ("shell") as a function of the magnetic fields applied in-plane ($B_\parallel$) and out-of-plane ($B_\perp$) to the BLG sheet. 
	At zero magnetic field, the four states are split into two Kramer's pairs, separated by $\Delta_\mathrm{SO}$. 
	Applying an out-of-plane magnetic field linearly shifts the energy of the states according to the spin and valley Zeeman effects  $E(B_\perp) = \frac{1}{2} 
	(\pm g_s \pm g_v) \mu_B B_{\perp}$,
	where $\mu_B$ is the Bohr magneton and $g_v$ is the valley $g$-factor, which quantifies the strength of the valley magnetic moment. Note that $g_v$ strongly depends on the QD wave functions and thus on the size of the QD \cite{Knothe2020Jun}. It is usually one order of magnitude larger than the spin $g$-factor, $g_s = 2$. 
	As the valley magnetic moment is oriented perpendicular to the BLG plane, in-plane $B$-fields only couple to the electron spin. However, as the SO coupling acts as an effective out-of-plane magnetic field close to the K-points, the spin states are polarized perpendicular the BLG plane (see insets in Fig. 1b)~\cite{Konschuh2012Mar}. Applying an in-plane magnetic field therefore shifts the states according to 
	$E(B_\parallel) = \pm \frac{1}{2} \sqrt{\Delta^2_\mathrm{SO} + (g_s \mu_B B_\parallel)^2}$
	recovering the linear spin Zeeman effect for high $B$-fields.

	\section{Results}
	\subsection{Device characterization}
\begin{figure}[!thb]
		\centering
		\includegraphics[draft=false,keepaspectratio=true,clip,width=0.6\linewidth]{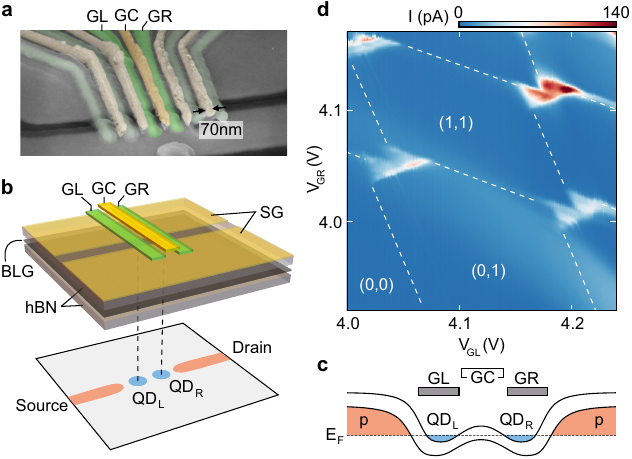}
		\caption[Fig02]{
			\textbf{a} False-color scanning electron micrograph of the metallic gates.
			A pair of split gates defines the conducting channel, which can be modulated by voltages applied to the finger gates. The gates used in the following are color coded. 
			\textbf{b} Schematic cross-section of the device. The upper part shows the metallic gates on top of the \new{hBN/BLG/hBN} van-der-Waals heterostructure. The lower part color codes  the charge carrier density within the channel and the two quantum dots (red: holes and blue: electrons). 
			\new{\textbf{c} Schematics of the band edge profile along the narrow channel, highlighting how the finger gates are used to form a DQD consisting of QD$_\text{L}$ and QD$_\text{R}$ connected to the p-type conducting channel.
				\textbf{d}} 
			Charge stability diagram showing the current through a double quantum dot as function of the potential applied to the two gate fingers, $V_\mathrm{GL}$ and $V_\mathrm{GR}$. A constant bias voltage of $V_\mathrm{b} = 1$~mV is applied and the central finger gate voltage is kept at $V_\mathrm{GC} = -4$~V.
		}
		\label{f2}
	\end{figure}

	\new{The devices consist} of a BLG flake, which has been encapsulated between two \new{($\approx25~$nm thick)} flakes of hexagonal boron nitride \new{(hBN)} and has been placed on a graphite flake, acting as a back gate (BG), using a dry van-der-Waals stacking technique. Cr/Au split gates (SGs) are deposited on top, forming a $2~\mu$m long and $130$~nm wide channel. Two layers of metallic Cr/Au finger gates (FGs) with a width of 70~nm and a pitch of 150~nm are fabricated across the channel. Details on the fabrication process can be found in Ref.~\cite{Banszerus2020Oct}. Fig.~\ref{f2}a shows a scanning electron micrograph of the gate structure, where the gates used as plunger gates are color coded. Fig.~\ref{f2}b shows a schematic cross section through the %van-der-Waals 
	heterostructure and the gate stack highlighting the formation of the QDs and source-drain regions by electrostatic soft-confinement. 
	All measurements are performed in a helium dilution refrigerator at a base temperature of $10$~mK, using standard DC measurement techniques.
	
	QDs are created using three layers of top gates, following previous studies of gate-defined BLG QDs~\cite{Banszerus2018Aug, Eich2018Aug, Eich2018Jul, Kurzmann2019Jul, Tong2021Jan, Banszerus2020Oct}. A band gap is opened by applying an out-of-plane displacement field~\cite{Zhang2009Jun,Oos2007Dec} with the help of the SG ($V_\mathrm{SG} = 1.73$~V) and BG ($V_\mathrm{BG} = -1.56$~V), and the Fermi energy ($E_\text{F}$) is tuned into the band gap. This leaves a narrow p-type conducting channel, connecting source and drain. A single electron DQD can be formed using adjacent FGs on the lower FG layer (GL and GR), locally overcompensating the BG voltage (see lower illustration in Fig.~2b \new{and the band edge diagram in Fig.~\ref{f2}c, highlighting the potential landscape along the narrow channel})~\cite{Banszerus2020Oct}. By applying $V_\mathrm{GC} = -4$~V to the central FG between GL and GR, the interdot tunnel coupling is reduced in order to enhance the energy resolution of the bias spectroscopy measurements. \new{Fig.~\ref{f2}d} shows a charge stability diagram of the first four pairs of triple points (see \new{supplementary information} for more details). %\new{Figure~\ref{f2}d displays a schematic representation of the band alignment, i.e. the potential landscape of the device along the channel.}
	
\begin{figure}[!thb]
		\centering
		\includegraphics[draft=false,keepaspectratio=true,clip,width=0.8\linewidth]{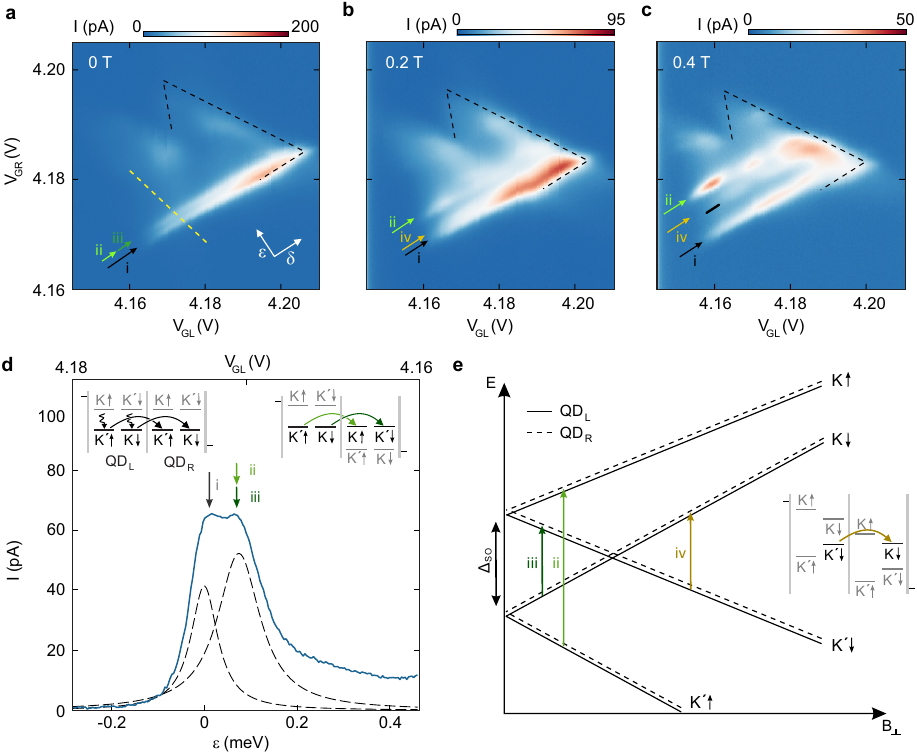}
		\caption[Fig03]{
			\textbf{a-c} Charge stability diagrams of the first pair of triple points ((1,0)-(0,1) transition) measured at $V_\mathrm{b} = 1~$mV and \textbf{a} $B_\perp=0$~T, \textbf{b} $B_\perp=0.2$~T, \textbf{c} $B_\perp=0.4$~T. 
			The white arrows in panel \textbf{a} indicate the orthogonal detuning axis $\new{\varepsilon}$ and the DQD's common energy $\delta$. At zero magnetic field two transition lines are visible (green and black arrows). 
			\textbf{d} Cut along the yellow dashed line in the first triple point at $B=0$~T. Two Lorentzian peaks are fit to the data to extract a splitting of $\Delta_\mathrm{SO} = 68\pm 7 \, \mu$eV.
			Inset: Schematic energy diagrams of a DQD in the finite bias regime for different interdot detuning energies $\new{\varepsilon}$, illustrating resonant transport through the ground state of each QD (transition (i); left inset) and resonant transport at $\new{\varepsilon} = \Delta_\mathrm{SO}$ (transitions (ii) and (iii); right inset). 
			\textbf{e} Schematic of the single particle energy spectra of the first orbital of each QD as a function of  $B_\perp$. The arrows (color coding as in panels \textbf{a-c}) indicate spin conserving transitions from single particle states in the left QD (solid lines) to single particle states in the right QD (dashed lines). 
			The inset shows the transition (iv) at finite magnetic field.
		}
		\label{f3}
	\end{figure}

	Next, we focus on the (0,1)-(1,0) charge transition, where each of the QDs is at most occupied by a single electron. Importantly, the combined tunneling rate is reduced to $\Gamma<1$~GHz by GC on the upmost gate layer (see \new{Figs.~\ref{f2}a-c}). This reduces the tunnel broadening of the resonance lines and strongly suppresses transport if the states in the two QDs are off resonance. Figs.~\ref{f3}a-c show finite bias charge stability diagrams of the first triple point pair at out-of-plane magnetic fields of \new{$B_\perp = 0$, $0.2$ and $0.4$~T}. In order to describe the configurations of the DQD, we introduce the orthogonal axes $\delta$ and $\new{\varepsilon}$, which describe how far the states in both QDs are tuned into the bias window ($\delta$) and how large their energy detuning ($\new{\varepsilon}$) is, respectively \footnote{In the single electron regime, the QD transitions (chemical potentials) are equivalent to the single particle energies.}. 
	At zero $B$-field, two resonances close to zero detuning (resonances (i) and (ii), (iii)) are visible, while the rest of the triple point shows only suppressed current. 
	Increasing $B_\perp$ shifts one of the resonances (ii) to higher detuning (compare green arrows in Figs.~\ref{f3}a-c). Eventually, a third resonance appears (iv), which does not extend as far on the $\delta$-axis as the other transitions (see Fig.~\ref{f3}c).

	The nature of these resonances can be explained in terms of transitions from single-particle states in the left QD (QD$_\mathrm{L}$) to single-particle states in the right QD (QD$_\mathrm{R}$). 
	For the present interdot tunneling times ($\approx 10$~ns~\footnote{We consider the combined tunneling rate, $\Gamma_\mathrm{comb}$ to be limited by the interdot tunneling rate, $\Gamma_\mathrm{m}$: $I/e=\Gamma_\mathrm{comb}\approx\Gamma_\mathrm{m}$, \new{where $I$ is the current through the DQD device and $e$ the elementary charge.}}), we assume that the electron spin is entirely conserved, while phonon assisted valley relaxation may occur on these time scales, as well as during interdot tunneling~\cite{Reichardt2016Jun}. \new{This is supported by the absence of any $\delta$ dependence of the transitions}. Fig.~\ref{f3}d shows a line cut through the triple point in Fig.~\ref{f3}a along the yellow dashed line. The two transitions (black arrow and green arrows in Fig.~\ref{f3}a) result in two distinct peaks in the tunneling current. The first resonance, (i) occurs at $\new{\varepsilon} = 0$, where every state in the left QD can tunnel into its equivalent state in the right QD, highlighted by the black arrow (see left schematic in Fig.~3d). The second resonance occurs at $\new{\varepsilon} = \Delta_\mathrm{SO} = 68\pm 7 \, \mu$eV, where two processes are possible, both requiring valley flips, namely transition (ii): $(K'\!\!\uparrow)_\mathrm{L} \,\, \Rightarrow \,\, (K\!\!\uparrow)_\mathrm{R}$ and transition (iii): $(K\!\!\downarrow)_\mathrm{L} \, \Rightarrow \, (K'\!\!\downarrow)_\mathrm{R}$, highlighted by the two green arrows (see right schematic in Fig.~3d). 
	
	\subsection{Magneto-transport spectroscopy}
	When applying an out-of-plane $B$-field, the energies of the single particle states shift according to their spin and valley Zeeman effect, as depicted on the right-hand side of Fig.~\ref{f1}b. Consequently, the detuning energy necessary for the transition %$(K^{-}\!\!\downarrow)_\mathrm{L} \,\, \Rightarrow \,\, (K^{+}\!\!\downarrow)_\mathrm{R}$
	(ii) increases linearly with magnetic field, highlighted by the light green arrow in Fig.~\ref{f3}e. The observed increase in detuning energy corresponds to a valley $g$-factor of $g_v \approx 15$. This observation validates the assumption that valley flips are allowed, since otherwise interdot transitions should not shift as function of $B_\perp$. The detuning required for transition (iii), decreases to zero, once the involved states are equal in energy, which is the case at $B_\perp \approx 0.18$~T (see Fig.~3e). At higher magnetic fields, the reversed process, transition (iv): $(K'\!\!\downarrow)_\mathrm{L} \,\, \Rightarrow \,\, (K\!\!\downarrow)_\mathrm{R}$ becomes possible, which also shifts with a valley $g$-factor of $g_v \approx 15$. This transition is marked by the orange arrow in Fig.~\ref{f3}c and offset to the transition (ii) by $\Delta \new{\varepsilon} = 2\Delta_\mathrm{SO}$, which becomes apparent from Fig.~3e. Since transition (iv) originates from an excited state (ES) in the QD$_\mathrm{L}$, it only becomes accessible as soon as the ES enters the bias window. Therefore, the transition line (iv) has a shorter extend along the $\delta$-axis, as it only sets in at finite $\delta$, which is highlighted by the black line in Fig.~3c. The observation of transition (iv) also justifies the assumption of spin conservation. If the spin lifetime would be shorter than the tunneling rate, $(K'\!\!\downarrow)_\mathrm{L}$ would decay into $(K'\!\!\uparrow)_\mathrm{L}$, blocking this process and the resonance would not be visible. Please note that the tunneling current corresponding to transitions (ii)-(iv) is energy ($\new{\varepsilon}$) and $\delta$-dependent as resonant tunneling (in particular with the drain reservoir) leads to higher tunneling currents (see Figs.~\ref{f3}a-c).

\begin{figure}[!thb]
		\centering
		\includegraphics[draft=false,keepaspectratio=true,clip,width=1\linewidth]{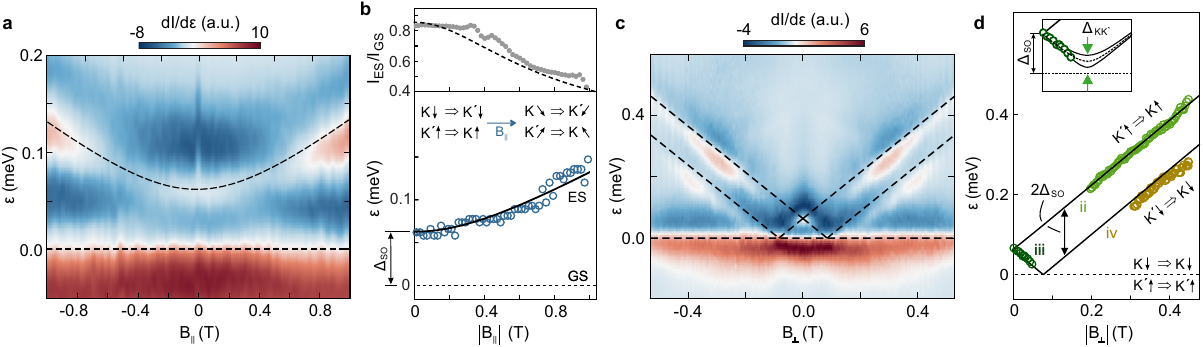}
		\caption[Fig04]{
			\textbf{a} 
			Magnetotransport measurements showing $dI/d\new{\varepsilon}$ as function of detuning energy and in-plane $B$-field. The horizontal dashed line marks the ground state (GS) transition ($\new{\varepsilon}=0$) and the curved line marks the excited state (ES) transition. 
			\textbf{b} 
			%The upper panel shows the ratio of the ES current with respect to the GS current. The dashed line describes the expected decay in amplitude of the ES due to the spins tilting in-plane and is given by $I_\mathrm{ES}/I_\mathrm{GS} \propto \sqrt{1 + (g_s \mu_B B_{\parallel} / \Delta_\mathrm{SO})^2}^{-1}$. The lower panel shows the extracted ES energy as function of $B_\parallel$. This data is used for fitting (see text). 
			The upper panel shows the ratio of the ES current with respect to the GS current. From a simple projection model we obtain the dashed line, which describes the expected decay in amplitude of the ES due to the spins tilting in-plane.
			\textbf{c} 
			Similar data as in panel \textbf{a} but for out-of-plane $B$-field. The dashed lines mark all transitions described in Fig.~\textbf{3e}.
			\textbf{d} 
			Extracted transition energies as function of $B_\perp$ highlighting the transitions (ii), (iii) and (iv) (see labels). The inset shows a low-energy close-up around the transition (iii) highlighting the presence of a finite $K-K'$ mixing. The dashed line corresponds to transition energies including a $\Delta_\mathrm{KK'} = 20 \,\mu$eV. The solid lines below and above correspond to $\Delta_\mathrm{KK'} = 10 \,\mu$eV and $30 \,\mu$eV, respectively.
		}
		\label{f4}
	\end{figure}

	Fig.~\ref{f4} shows the interdot transitions as function of in-plane (Figs.~4a,b) and out-of-plane (Figs.~4c,d) $B$-field highlighting the spin and valley texture of the low energy spectrum. In Fig.~4a we show 
	the derivative of the tunneling current $I$ with respect to $\new{\varepsilon}$ as function of energy and $B_\parallel$. Apart from the zero detuning transition (horizontal dashed line), one additional feature is visible (curved dashed line), which corresponds to the transitions from the energetically lower Kramer's pair to the energetically higher Kramer's pair, namely $(K'\!\!\uparrow)_\mathrm{L} \, \Rightarrow \, (K\!\!\uparrow)_\mathrm{R}$ and $(K\!\!\downarrow)_\mathrm{L} \,\, \Rightarrow  (K'\!\!\downarrow)_\mathrm{R}$, as highlighted by the schematic insets of Fig.~3d.
	Increasing the in-plane $B$-field increases the energy difference between the Kramer's pairs due to the spin Zeeman effect and therefore the required detuning shifts according to 
	%\begin{equation}
	%    \new{\varepsilon} =  \sqrt{\Delta_\mathrm{SO}^2 + (g_s \mu_B  B)^2}. 
	%    \label{eq:inplane_fit}
	%\end{equation}
	%
	$    \new{\varepsilon} =  \sqrt{\Delta_\mathrm{SO}^2 + (g_s \mu_B  B)^2}.$ 
	
	Fitting this equation with a fixed $g_s = 2$, which is in good agreement with earlier measurements~\cite{Banszerus2018Aug}, incl. electron spin resonance experiments~\cite{Lyon2017Aug}, yields $\Delta_\mathrm{SO} = 62 \pm 6 ~ \mu$eV.
	This results in the dashed and solid line in Figs.~4a,b showing
	good agreement with the experiment.
	Applying an in-plane magnetic field tilts the spin from an out-of-plane orientation induced by the SO coupling into the plane of the BLG (see insets in Fig.~\ref{f4}b). This effect continuously reduces the overlap between the spin states from different Kramer's pairs, e.g. the spin state from $K'\!\!\uparrow$ is not perfectly parallel to $K\!\!\uparrow$ anymore, until for large $B_\parallel$ they will eventually be completely orthogonal. 
	This effect becomes visible in the tunneling current of the ES transition, which decreases with increasing in-plane magnetic field.
	In the top panel of Fig.~4b, we show the ratio between the current through the exited and the ground state ($\new{\varepsilon}=0$) as a function of $B_\parallel$ highlighting this effect, which is in  good agreement with what is expected from theory (see dashed line and figure caption).
	
	Fig.~\ref{f4}c shows the derivative of the tunneling current with respect to $\new{\varepsilon}$, as function of energy and $B_\perp$. 
	Here, according to Fig.~1b the transition spectra is significantly richer and all three transitions (ii), (iii) and (iv) discussed in Fig.~3e can be observed (see dashed lines and labels in Fig.~4d). 
	If we denote the valley $g$-factor in the left and right QD with $g_{v,\mathrm{L}}$ and $g_{v,\mathrm{R}}$, respectively, then we can express the different transition energies as a function of $B_{\perp}$ by 
	%\begin{equation}
	$\new{\varepsilon}_\mathrm{ii} = \Delta_\mathrm{SO} + \frac{1}{2} (g_{v,\mathrm{R}} + \, g_{v,\mathrm{L}}) \mu_B  B_{\perp}$, 
	%\end{equation} 
	%\begin{equation}
	$\new{\varepsilon}_\mathrm{iii} = \Delta_\mathrm{SO} - \frac{1}{2} (g_{v,\mathrm{R}} + \, g_{v,\mathrm{L}}) \mu_B B_{\perp}$, 
	%\end{equation}
	%\begin{equation}
	$\new{\varepsilon}_\mathrm{iv} = -\Delta_\mathrm{SO} + \frac{1}{2} (g_{v,\mathrm{R}} + \, g_{v,\mathrm{L}}) \mu_B  B_{\perp}$.
	%\end{equation}
	%
	Here, the electron spin Zeeman effect has no influence on the transition energies, as transitions only occur between states of the same electron spin. 
	Taking $\Delta_\mathrm{SO}$ from the analysis of the in-plane $B$-field data (Figs.~4a,b) and choosing the valley g-factors to be $g_{v,\mathrm{L}} = g_{v,\mathrm{R}} = 15$
	we find -- without any additional parameter -- good agreement with the experimental data (see dashed and solid lines in Figs.~4c,d).
	The values of $g_v$ are similar to those reported in previous studies of similar BLG QDs and compatible with theoretical calculations~\cite{Banszerus2020Oct,Tong2021Jan}.
	Considering the same geometry of both QDs and the similar voltages applied to GL and GR, it is reasonable to assume that both QDs have very similar valley $g$-factors. 
	Note that as soon as a transition requires negative detuning, it becomes Coulomb \new{blockaded}, which is the reason why no transition is observed below $\new{\varepsilon} = 0$. 
	The lack of clear signatures of the transition (ii) at low $B$-field (see Figs.~4c,d) can be explained by their reduced tunneling current compared to transition (iii) due to the stronger detuning of the $(K'\!\!\uparrow)_\mathrm{L}$ ground state from the source chemical potential compared to $(K\!\!\downarrow)_\mathrm{L}$, combined with strong resonant tunneling from source to the left QD. 
	
	By closely inspecting the transition (iii), especially  close to the $B$-field regime where the $K\!\!\downarrow$ state is crossing the $K'\!\!\downarrow$ state (i.e. around $\new{\varepsilon} = 0$), we can provide an estimate of the upper limit of a possible disorder-induced mixing of the $K$ and $K'$ states.
	The inset of Fig.~4d shows a close up, where we included the expected transition energies for different values of $\Delta_\mathrm{KK'}$ \new{(see also supplementary information)}.
	We do not observe any anticrossing \new{within the margin of the energy resolution of our measurement, neither for the device presented in Fig.~4d nor for a second single-electron DQD device presented in the supplementary information.} From this comparison we estimate that $\Delta_\mathrm{KK'}$ is surely not exceeding a value of $20 \, \mu$eV in \new{both DQD devices}. Note that this upper limit is significantly smaller than for carbon nanotubes with values on the order of 100~$\mu$eV~\cite{Kuemmeth2008Mar,Jespersen2011Apr}.
	\new{In carbon nanotubes the present $\Delta_\mathrm{KK'}$ may also influence the magnitude of the zero-field splitting of the Kramer's pairs and thus potentially lead to an overestimation of $\Delta_\mathrm{SO}$.
		As the zero field splitting consists of the quadratic sum of the two effects, $\sqrt{\Delta_\mathrm{SO}^2 + \Delta_\mathrm{KK'}^2}$, and our observed splitting is at least a factor three larger than $\Delta_\mathrm{KK'}$, the influence of the intervalley mixing on the observed values of $\Delta_\mathrm{SO}$ is smaller than 10$\%$ and thus lies within the range of our measurement uncertainties.
		%The zero-field splitting comprises of the quadratic sum of the two effects: $\sqrt{\Delta_\mathrm{SO}^2 + \Delta_\mathrm{KK'}^2}$. Since the observed zero-field splitting is at least a factor of three to four larger than $\Delta_\mathrm{KK'}$, the impact of intervalley mixing on the observed values of $\Delta_\mathrm{SO}$ is smaller than 10$\%$ and thus within the margin of measurement uncertainties.
	}\\ 
	\section{Discussion}
	\new{
		Apart from the Kane-Mele SO coupling, which is intrinsically present in graphene and BLG, extrinsic Bychkov-Rashba SO coupling and pseudospin inversion asymmetry (or principal plane asymmetry) SO coupling can in principle also play a role~\cite{Konschuh2012Mar,Kane2005Nov,Min2006Oct,Yao2007Jan,Zollner2019Mar,Prada2021Mar,Gmitra2013Jun}.
		The latter, which arises for example when placing graphene or BLG on substrates, such as e.g. hBN, depends on the magnitude of $k$ (measured from the corners of the Brillouin zone) and is thus suppressed at the $K$ and $K'$-points~\cite{Zollner2019Mar}, where our devices are operated.
		The Rashba-type SO coupling needs to be discussed in more detail, since in our devices the inversion symmetry is explicitly broken by the applied out-of-plane displacement field.
		%
		%While the latter arises when placing graphene or BLG on substrates, such as e.g. hBN, the term is strongly suppressed close to $K$ and $K'$-points~\cite{Zollner2019Mar}, %~\footnote{Private communication with J. Fabian}, 
		%where our devices are operated. 
		%
		%Rashba type SO coupling has to be discussed, since in our BLG devices the inversion symmetry is explicitly broken by the applied out-of-plane displacement field.
		%Apart from the Kane-Mele SO gap, which is intrinsically present in graphene and BLG, Bychkov-Rashba type SO coupling and principal may in principle play a role as well, since the inversion symmetry (which forbids Rashba SO coupling) is in our BLG devices explicitly broken by applying a perpendicular electric displacement field.
		%
		The breaking of inversion symmetry and the magnitude of the resulting Bychkov-Rashba SO gaps have theoretically been investigated by Kane and Mele for single-layer graphene~\cite{Kane2005Nov} and by Konschuh et al. for BLG~\cite{Konschuh2012Mar} . 
		Both studies conclude that the Rashba type SO coupling is negligible ($\approx1~\mu$eV) compared to the Kane-Mele coupling term. 
		In addition, the Bychkov-Rashba SO coupling term is expected to be strongly suppressed in BLG single-electron QDs, since specifically in BLG this term vanishes for the low energy bands close to the $K$ and $K'$-points~\cite{Konschuh2012Mar}.
		%, where the wavefunction in in such effective zero-dimensional systems is confined to.
		%
		Thus, all this is expected to lead to a displacement-field-independent SO gap that can be experimentally verified.
		%Thus, it is expected that all these leads to a displacement field independent SO gap, which can be tested experimentally.
		%
		\\
		In Fig.~5 we show the SO gap as function of displacement field, $D$. Here, $\Delta_\mathrm{SO}$ has been extracted from zero $B$-field data similar to the measurements shown in Figs.~3a,d but with different back and split gate voltages such that the $D$-field is tuned from $D=0.24~$V/nm to $0.34~$V/nm, also resulting in different band gaps in the BLG as highlighted by the insets in Fig.~5.
		From all data presented in Fig.~5 -- including data from a second single-electron DQD device (red triangles) and data from a different single-electron QD device (yellow square, more details in the supplementary information) -- we conclude that the observed SO gaps are all consistent and within the error bars constant over the investigated $D$-field range, with a mean value around $\Delta_\mathrm{SO}\approx60~\mu$eV. 
		%
		%To verify these considerations experimentally, we tune the displacement field in the device between $D=0.24~$V/nm and $D=0.34~$V/nm by alternating the back and split gate voltages and measure the zero field splitting from similar measurements as shown in Fig.~3a. In Figure~5, $\Delta_\mathrm{SO}$, as function of $D$ is plotted. 
		%
		%From the data it can be seen that $\Delta_\mathrm{SO}\approx60~\mu$eV remains constant within the measurement resolution. The blue data points corresponds to $\Delta_\mathrm{SO}$ measured in a second device (see supplementary information), which is very close to the data obtained from the first device. 
		%
		From the absence of any dependency of $\Delta_\mathrm{SO}$ as function of the strength of the potential breaking the inversion symmetry, we conclude that the experimentally extracted $\Delta_\mathrm{SO}$ is dominated by the Kane-Mele coupling term.
		Interestingly, our $\Delta_\mathrm{SO}$ values are slightly larger than what has been extracted in previous experiments performed in bulk graphene on trenched SiO$_2$~\cite{Sichau2019Feb}. 
		Also this value is larger than theory predicted~\cite{Konschuh2012Mar}, but might be explained by an enhancement due to phonon-assisted SO coupling~\cite{Ochoa2012Dec}.
		In our case we expect that the SO coupling is slightly enhanced due to the proximity effect when encapsulating BLG with hBN crystals~\cite{Banszerus2020May}, very similar to the proximity enhanced SO coupling when placing BLG on WSe$_2$~\cite{Island2019Jul}.
		%proximity enhanced SOC by the hBN crystals on top and below the BLG~\cite{Banszerus2020May} or even by a potential coupling of the spin and valley magnetic moments close to K and K'.
	}

	In summary, we studied the low-energy excited state spectrum of a gate-defined single-electron quantum dot in bilayer graphene. We find a spin-valley coupling dominated by a Kane-Mele type SO coupling with \new{$\Delta_\mathrm{SO}\approx60~\mu$eV}, giving rise to two Kramer's pairs with either parallel or anti-parallel spin-valley orientation.
	The  small value for  $\Delta_\mathrm{KK'}$ ($<20 \, \mu$eV) is not entirely unexpected for flat and disorder-free BLG, and raises the hope that the existing spin-valley coupling -- without $K - K'$ mixing  -- will be helpful for a future qubit operation.\\

	\textbf{Acknowledgements} \\
	The authors \new{acknowledge discussions with J. Fabian, A. Knothe and V. Fal'ko and} thank J. Klos for help with the SEM micrographs.
	This project has received funding from the European Union's Horizon 2020 research and innovation programme under grant agreement No. 881603 (Graphene Flagship) and from the European Research Council (ERC) under grant agreement No. 820254, the Deutsche Forschungsgemeinschaft (DFG, German Research Foundation) under Germany's Excellence Strategy - Cluster of Excellence Matter and Light for Quantum Computing (ML4Q) EXC 2004/1 - 390534769, through DFG (STA 1146/11-1), and by the Helmholtz Nano Facility~\cite{Albrecht2017May}. Growth of hexagonal boron nitride crystals was supported by the Elemental Strategy Initiative conducted by the MEXT, Japan, Grant Number JPMXP0112101001,  JSPS KAKENHI Grant Numbers JP20H00354 and the CREST(JPMJCR15F3), JST.\\\\
	\textbf{Author contributions}\\
	L.B., S.M., E.I, S.T. and F.L. fabricated the device, L.B., S.M., C.Ste. and C.V. performed the measurements and analyzed the data. K.W. and  T.T.  synthesized  the  hBN  crystals. C.V. and C.Sta. supervised the project. L.B., S.M., C.Ste., C.V. and C.Sta. wrote the manuscript with contributions from all authors. L.B. and S.M contributed equally to this work.\\\\
	\textbf{Data availability}\\
	The data supporting the findings of this study will be made available by the authors upon reasonable request. \\\\
	\textbf{Competing interests}\\
	The authors declare no competing interests.
	
	\bibliography{Literature}

\begin{thebibliography}{34}%
\makeatletter
\providecommand \@ifxundefined [1]{%
 \@ifx{#1\undefined}
}%
\providecommand \@ifnum [1]{%
 \ifnum #1\expandafter \@firstoftwo
 \else \expandafter \@secondoftwo
 \fi
}%
\providecommand \@ifx [1]{%
 \ifx #1\expandafter \@firstoftwo
 \else \expandafter \@secondoftwo
 \fi
}%
\providecommand \natexlab [1]{#1}%
\providecommand \enquote  [1]{``#1''}%
\providecommand \bibnamefont  [1]{#1}%
\providecommand \bibfnamefont [1]{#1}%
\providecommand \citenamefont [1]{#1}%
\providecommand \href@noop [0]{\@secondoftwo}%
\providecommand \href [0]{\begingroup \@sanitize@url \@href}%
\providecommand \@href[1]{\@@startlink{#1}\@@href}%
\providecommand \@@href[1]{\endgroup#1\@@endlink}%
\providecommand \@sanitize@url [0]{\catcode `\\12\catcode `\$12\catcode
  `\&12\catcode `\#12\catcode `\^12\catcode `\_12\catcode `\%12\relax}%
\providecommand \@@startlink[1]{}%
\providecommand \@@endlink[0]{}%
\providecommand \url  [0]{\begingroup\@sanitize@url \@url }%
\providecommand \@url [1]{\endgroup\@href {#1}{\urlprefix }}%
\providecommand \urlprefix  [0]{URL }%
\providecommand \Eprint [0]{\href }%
\providecommand \doibase [0]{http://dx.doi.org/}%
\providecommand \selectlanguage [0]{\@gobble}%
\providecommand \bibinfo  [0]{\@secondoftwo}%
\providecommand \bibfield  [0]{\@secondoftwo}%
\providecommand \translation [1]{[#1]}%
\providecommand \BibitemOpen [0]{}%
\providecommand \bibitemStop [0]{}%
\providecommand \bibitemNoStop [0]{.\EOS\space}%
\providecommand \EOS [0]{\spacefactor3000\relax}%
\providecommand \BibitemShut  [1]{\csname bibitem#1\endcsname}%
\let\auto@bib@innerbib\@empty
%</preamble>
\bibitem [{\citenamefont {Trauzettel}\ \emph {et~al.}(2007)\citenamefont
  {Trauzettel}, \citenamefont {Bulaev}, \citenamefont {Loss},\ and\
  \citenamefont {Burkard}}]{Trauzettel2007Mar}%
  \BibitemOpen
  \bibfield  {author} {\bibinfo {author} {\bibfnamefont
  {Bj{\ifmmode\ddot{o}\else\"{o}\fi}rn}\ \bibnamefont {Trauzettel}}, \bibinfo
  {author} {\bibfnamefont {Denis~V.}\ \bibnamefont {Bulaev}}, \bibinfo {author}
  {\bibfnamefont {Daniel}\ \bibnamefont {Loss}}, \ and\ \bibinfo {author}
  {\bibfnamefont {Guido}\ \bibnamefont {Burkard}},\ }\bibfield  {title}
  {\enquote {\bibinfo {title} {{Spin qubits in graphene quantum dots}},}\
  }\href {\doibase 10.1038/nphys544} {\bibfield  {journal} {\bibinfo  {journal}
  {Nat. Phys.}\ }\textbf {\bibinfo {volume} {3}},\ \bibinfo {pages} {192--196}
  (\bibinfo {year} {2007})}\BibitemShut {NoStop}%
\bibitem [{\citenamefont {Wu}\ \emph {et~al.}(2013)\citenamefont {Wu},
  \citenamefont {Lue},\ and\ \citenamefont {Chen}}]{Wu2013Sep}%
  \BibitemOpen
  \bibfield  {author} {\bibinfo {author} {\bibfnamefont {G.~Y.}\ \bibnamefont
  {Wu}}, \bibinfo {author} {\bibfnamefont {N.-Y.}\ \bibnamefont {Lue}}, \ and\
  \bibinfo {author} {\bibfnamefont {Y.-C.}\ \bibnamefont {Chen}},\ }\bibfield
  {title} {\enquote {\bibinfo {title} {{Quantum manipulation of valleys in
  bilayer graphene}},}\ }\href {\doibase 10.1103/PhysRevB.88.125422} {\bibfield
   {journal} {\bibinfo  {journal} {Phys. Rev. B}\ }\textbf {\bibinfo {volume}
  {88}},\ \bibinfo {pages} {125422} (\bibinfo {year} {2013})}\BibitemShut
  {NoStop}%
\bibitem [{\citenamefont {McCann}\ and\ \citenamefont
  {Koshino}(2013)}]{McCann2013Apr}%
  \BibitemOpen
  \bibfield  {author} {\bibinfo {author} {\bibfnamefont {Edward}\ \bibnamefont
  {McCann}}\ and\ \bibinfo {author} {\bibfnamefont {Mikito}\ \bibnamefont
  {Koshino}},\ }\bibfield  {title} {\enquote {\bibinfo {title} {{The electronic
  properties of bilayer graphene}},}\ }\href {\doibase
  10.1088/0034-4885/76/5/056503} {\bibfield  {journal} {\bibinfo  {journal}
  {Rep. Prog. Phys.}\ }\textbf {\bibinfo {volume} {76}},\ \bibinfo {pages}
  {056503} (\bibinfo {year} {2013})}\BibitemShut {NoStop}%
\bibitem [{\citenamefont {Knothe}\ and\ \citenamefont
  {Fal'ko}(2018)}]{Knothe2018Oct}%
  \BibitemOpen
  \bibfield  {author} {\bibinfo {author} {\bibfnamefont {Angelika}\
  \bibnamefont {Knothe}}\ and\ \bibinfo {author} {\bibfnamefont {Vladimir}\
  \bibnamefont {Fal'ko}},\ }\bibfield  {title} {\enquote {\bibinfo {title}
  {{Influence of minivalleys and Berry curvature on electrostatically induced
  quantum wires in gapped bilayer graphene}},}\ }\href {\doibase
  10.1103/PhysRevB.98.155435} {\bibfield  {journal} {\bibinfo  {journal} {Phys.
  Rev. B}\ }\textbf {\bibinfo {volume} {98}},\ \bibinfo {pages} {155435}
  (\bibinfo {year} {2018})}\BibitemShut {NoStop}%
\bibitem [{\citenamefont {Eich}\ \emph
  {et~al.}(2018{\natexlab{a}})\citenamefont {Eich}, \citenamefont {Pisoni},
  \citenamefont {Pally}, \citenamefont {Overweg}, \citenamefont {Kurzmann},
  \citenamefont {Lee}, \citenamefont {Rickhaus}, \citenamefont {Watanabe},
  \citenamefont {Taniguchi}, \citenamefont {Ensslin},\ and\ \citenamefont
  {Ihn}}]{Eich2018Aug}%
  \BibitemOpen
  \bibfield  {author} {\bibinfo {author} {\bibfnamefont {Marius}\ \bibnamefont
  {Eich}}, \bibinfo {author} {\bibfnamefont {Riccardo}\ \bibnamefont {Pisoni}},
  \bibinfo {author} {\bibfnamefont {Alessia}\ \bibnamefont {Pally}}, \bibinfo
  {author} {\bibfnamefont {Hiske}\ \bibnamefont {Overweg}}, \bibinfo {author}
  {\bibfnamefont {Annika}\ \bibnamefont {Kurzmann}}, \bibinfo {author}
  {\bibfnamefont {Yongjin}\ \bibnamefont {Lee}}, \bibinfo {author}
  {\bibfnamefont {Peter}\ \bibnamefont {Rickhaus}}, \bibinfo {author}
  {\bibfnamefont {Kenji}\ \bibnamefont {Watanabe}}, \bibinfo {author}
  {\bibfnamefont {Takashi}\ \bibnamefont {Taniguchi}}, \bibinfo {author}
  {\bibfnamefont {Klaus}\ \bibnamefont {Ensslin}}, \ and\ \bibinfo {author}
  {\bibfnamefont {Thomas}\ \bibnamefont {Ihn}},\ }\bibfield  {title} {\enquote
  {\bibinfo {title} {{Coupled Quantum Dots in Bilayer Graphene}},}\ }\href
  {\doibase 10.1021/acs.nanolett.8b01859} {\bibfield  {journal} {\bibinfo
  {journal} {Nano Lett.}\ }\textbf {\bibinfo {volume} {18}},\ \bibinfo {pages}
  {5042--5048} (\bibinfo {year} {2018}{\natexlab{a}})}\BibitemShut {NoStop}%
\bibitem [{\citenamefont {Banszerus}\ \emph
  {et~al.}(2020{\natexlab{a}})\citenamefont {Banszerus}, \citenamefont
  {Rothstein}, \citenamefont {Fabian}, \citenamefont
  {M{\ifmmode\ddot{o}\else\"{o}\fi}ller}, \citenamefont {Icking}, \citenamefont
  {Trellenkamp}, \citenamefont {Lentz}, \citenamefont {Neumaier}, \citenamefont
  {Watanabe}, \citenamefont {Taniguchi}, \citenamefont {Libisch}, \citenamefont
  {Volk},\ and\ \citenamefont {Stampfer}}]{Banszerus2020Oct}%
  \BibitemOpen
  \bibfield  {author} {\bibinfo {author} {\bibfnamefont {L.}~\bibnamefont
  {Banszerus}}, \bibinfo {author} {\bibfnamefont {A.}~\bibnamefont
  {Rothstein}}, \bibinfo {author} {\bibfnamefont {T.}~\bibnamefont {Fabian}},
  \bibinfo {author} {\bibfnamefont {S.}~\bibnamefont
  {M{\ifmmode\ddot{o}\else\"{o}\fi}ller}}, \bibinfo {author} {\bibfnamefont
  {E.}~\bibnamefont {Icking}}, \bibinfo {author} {\bibfnamefont
  {S.}~\bibnamefont {Trellenkamp}}, \bibinfo {author} {\bibfnamefont
  {F.}~\bibnamefont {Lentz}}, \bibinfo {author} {\bibfnamefont
  {D.}~\bibnamefont {Neumaier}}, \bibinfo {author} {\bibfnamefont
  {K.}~\bibnamefont {Watanabe}}, \bibinfo {author} {\bibfnamefont
  {T.}~\bibnamefont {Taniguchi}}, \bibinfo {author} {\bibfnamefont
  {F.}~\bibnamefont {Libisch}}, \bibinfo {author} {\bibfnamefont
  {C.}~\bibnamefont {Volk}}, \ and\ \bibinfo {author} {\bibfnamefont
  {C.}~\bibnamefont {Stampfer}},\ }\bibfield  {title} {\enquote {\bibinfo
  {title} {{Electron{\textendash}Hole Crossover in Gate-Controlled Bilayer
  Graphene Quantum Dots}},}\ }\href {\doibase 10.1021/acs.nanolett.0c03227}
  {\bibfield  {journal} {\bibinfo  {journal} {Nano Lett.}\ }\textbf {\bibinfo
  {volume} {20}},\ \bibinfo {pages} {7709--7715} (\bibinfo {year}
  {2020}{\natexlab{a}})}\BibitemShut {NoStop}%
\bibitem [{\citenamefont {Lee}\ \emph {et~al.}(2020)\citenamefont {Lee},
  \citenamefont {Knothe}, \citenamefont {Overweg}, \citenamefont {Eich},
  \citenamefont {Gold}, \citenamefont {Kurzmann}, \citenamefont {Klasovika},
  \citenamefont {Taniguchi}, \citenamefont {Wantanabe}, \citenamefont
  {Fal{'}ko}, \citenamefont {Ihn}, \citenamefont {Ensslin},\ and\ \citenamefont
  {Rickhaus}}]{Lee2020Mar}%
  \BibitemOpen
  \bibfield  {author} {\bibinfo {author} {\bibfnamefont {Yongjin}\ \bibnamefont
  {Lee}}, \bibinfo {author} {\bibfnamefont {Angelika}\ \bibnamefont {Knothe}},
  \bibinfo {author} {\bibfnamefont {Hiske}\ \bibnamefont {Overweg}}, \bibinfo
  {author} {\bibfnamefont {Marius}\ \bibnamefont {Eich}}, \bibinfo {author}
  {\bibfnamefont {Carolin}\ \bibnamefont {Gold}}, \bibinfo {author}
  {\bibfnamefont {Annika}\ \bibnamefont {Kurzmann}}, \bibinfo {author}
  {\bibfnamefont {Veronika}\ \bibnamefont {Klasovika}}, \bibinfo {author}
  {\bibfnamefont {Takashi}\ \bibnamefont {Taniguchi}}, \bibinfo {author}
  {\bibfnamefont {Kenji}\ \bibnamefont {Wantanabe}}, \bibinfo {author}
  {\bibfnamefont {Vladimir}\ \bibnamefont {Fal{'}ko}}, \bibinfo {author}
  {\bibfnamefont {Thomas}\ \bibnamefont {Ihn}}, \bibinfo {author}
  {\bibfnamefont {Klaus}\ \bibnamefont {Ensslin}}, \ and\ \bibinfo {author}
  {\bibfnamefont {Peter}\ \bibnamefont {Rickhaus}},\ }\bibfield  {title}
  {\enquote {\bibinfo {title} {{Tunable Valley Splitting due to Topological
  Orbital Magnetic Moment in Bilayer Graphene Quantum Point Contacts}},}\
  }\href {\doibase 10.1103/PhysRevLett.124.126802} {\bibfield  {journal}
  {\bibinfo  {journal} {Phys. Rev. Lett.}\ }\textbf {\bibinfo {volume} {124}},\
  \bibinfo {pages} {126802} (\bibinfo {year} {2020})}\BibitemShut {NoStop}%
\bibitem [{\citenamefont {Garreis}\ \emph {et~al.}(2020)\citenamefont
  {Garreis}, \citenamefont {Knothe}, \citenamefont {Tong}, \citenamefont
  {Eich}, \citenamefont {Gold}, \citenamefont {Watanabe}, \citenamefont
  {Taniguchi}, \citenamefont {Fal'ko}, \citenamefont {Ihn}, \citenamefont
  {Ensslin},\ and\ \citenamefont {Kurzmann}}]{Garreis2020Nov}%
  \BibitemOpen
  \bibfield  {author} {\bibinfo {author} {\bibfnamefont {Rebekka}\ \bibnamefont
  {Garreis}}, \bibinfo {author} {\bibfnamefont {Angelika}\ \bibnamefont
  {Knothe}}, \bibinfo {author} {\bibfnamefont {Chuyao}\ \bibnamefont {Tong}},
  \bibinfo {author} {\bibfnamefont {Marius}\ \bibnamefont {Eich}}, \bibinfo
  {author} {\bibfnamefont {Carolin}\ \bibnamefont {Gold}}, \bibinfo {author}
  {\bibfnamefont {Kenji}\ \bibnamefont {Watanabe}}, \bibinfo {author}
  {\bibfnamefont {Takashi}\ \bibnamefont {Taniguchi}}, \bibinfo {author}
  {\bibfnamefont {Vladimir}\ \bibnamefont {Fal'ko}}, \bibinfo {author}
  {\bibfnamefont {Thomas}\ \bibnamefont {Ihn}}, \bibinfo {author}
  {\bibfnamefont {Klaus}\ \bibnamefont {Ensslin}}, \ and\ \bibinfo {author}
  {\bibfnamefont {Annika}\ \bibnamefont {Kurzmann}},\ }\bibfield  {title}
  {\enquote {\bibinfo {title} {{Shell Filling and Trigonal Warping in Graphene
  Quantum Dots}},}\ }\href {https://arxiv.org/abs/2011.07951v1} {\bibfield
  {journal} {\bibinfo  {journal} {arXiv}\ } (\bibinfo {year} {2020})},\ \Eprint
  {http://arxiv.org/abs/2011.07951} {2011.07951} \BibitemShut {NoStop}%
\bibitem [{\citenamefont {Eich}\ \emph
  {et~al.}(2018{\natexlab{b}})\citenamefont {Eich}, \citenamefont {Pisoni},
  \citenamefont {Overweg}, \citenamefont {Kurzmann}, \citenamefont {Lee},
  \citenamefont {Rickhaus}, \citenamefont {Ihn}, \citenamefont {Ensslin},
  \citenamefont {Herman}, \citenamefont {Sigrist}, \citenamefont {Watanabe},\
  and\ \citenamefont {Taniguchi}}]{Eich2018Jul}%
  \BibitemOpen
  \bibfield  {author} {\bibinfo {author} {\bibfnamefont {Marius}\ \bibnamefont
  {Eich}}, \bibinfo {author} {\bibfnamefont {Riccardo}\ \bibnamefont {Pisoni}},
  \bibinfo {author} {\bibfnamefont {Hiske}\ \bibnamefont {Overweg}}, \bibinfo
  {author} {\bibfnamefont {Annika}\ \bibnamefont {Kurzmann}}, \bibinfo {author}
  {\bibfnamefont {Yongjin}\ \bibnamefont {Lee}}, \bibinfo {author}
  {\bibfnamefont {Peter}\ \bibnamefont {Rickhaus}}, \bibinfo {author}
  {\bibfnamefont {Thomas}\ \bibnamefont {Ihn}}, \bibinfo {author}
  {\bibfnamefont {Klaus}\ \bibnamefont {Ensslin}}, \bibinfo {author}
  {\bibfnamefont {Franti{\ifmmode\check{s}\else\v{s}\fi}ek}\ \bibnamefont
  {Herman}}, \bibinfo {author} {\bibfnamefont {Manfred}\ \bibnamefont
  {Sigrist}}, \bibinfo {author} {\bibfnamefont {Kenji}\ \bibnamefont
  {Watanabe}}, \ and\ \bibinfo {author} {\bibfnamefont {Takashi}\ \bibnamefont
  {Taniguchi}},\ }\bibfield  {title} {\enquote {\bibinfo {title} {{Spin and
  Valley States in Gate-Defined Bilayer Graphene Quantum Dots}},}\ }\href
  {\doibase 10.1103/PhysRevX.8.031023} {\bibfield  {journal} {\bibinfo
  {journal} {Phys. Rev. X}\ }\textbf {\bibinfo {volume} {8}},\ \bibinfo {pages}
  {031023} (\bibinfo {year} {2018}{\natexlab{b}})}\BibitemShut {NoStop}%
\bibitem [{\citenamefont {Kurzmann}\ \emph {et~al.}(2019)\citenamefont
  {Kurzmann}, \citenamefont {Eich}, \citenamefont {Overweg}, \citenamefont
  {Mangold}, \citenamefont {Herman}, \citenamefont {Rickhaus}, \citenamefont
  {Pisoni}, \citenamefont {Lee}, \citenamefont {Garreis}, \citenamefont {Tong},
  \citenamefont {Watanabe}, \citenamefont {Taniguchi}, \citenamefont
  {Ensslin},\ and\ \citenamefont {Ihn}}]{Kurzmann2019Jul}%
  \BibitemOpen
  \bibfield  {author} {\bibinfo {author} {\bibfnamefont {A.}~\bibnamefont
  {Kurzmann}}, \bibinfo {author} {\bibfnamefont {M.}~\bibnamefont {Eich}},
  \bibinfo {author} {\bibfnamefont {H.}~\bibnamefont {Overweg}}, \bibinfo
  {author} {\bibfnamefont {M.}~\bibnamefont {Mangold}}, \bibinfo {author}
  {\bibfnamefont {F.}~\bibnamefont {Herman}}, \bibinfo {author} {\bibfnamefont
  {P.}~\bibnamefont {Rickhaus}}, \bibinfo {author} {\bibfnamefont
  {R.}~\bibnamefont {Pisoni}}, \bibinfo {author} {\bibfnamefont
  {Y.}~\bibnamefont {Lee}}, \bibinfo {author} {\bibfnamefont {R.}~\bibnamefont
  {Garreis}}, \bibinfo {author} {\bibfnamefont {C.}~\bibnamefont {Tong}},
  \bibinfo {author} {\bibfnamefont {K.}~\bibnamefont {Watanabe}}, \bibinfo
  {author} {\bibfnamefont {T.}~\bibnamefont {Taniguchi}}, \bibinfo {author}
  {\bibfnamefont {K.}~\bibnamefont {Ensslin}}, \ and\ \bibinfo {author}
  {\bibfnamefont {T.}~\bibnamefont {Ihn}},\ }\bibfield  {title} {\enquote
  {\bibinfo {title} {{Excited States in Bilayer Graphene Quantum Dots}},}\
  }\href {\doibase 10.1103/PhysRevLett.123.026803} {\bibfield  {journal}
  {\bibinfo  {journal} {Phys. Rev. Lett.}\ }\textbf {\bibinfo {volume} {123}},\
  \bibinfo {pages} {026803} (\bibinfo {year} {2019})}\BibitemShut {NoStop}%
\bibitem [{\citenamefont {Knothe}\ and\ \citenamefont
  {Fal'ko}(2020)}]{Knothe2020Jun}%
  \BibitemOpen
  \bibfield  {author} {\bibinfo {author} {\bibfnamefont {Angelika}\
  \bibnamefont {Knothe}}\ and\ \bibinfo {author} {\bibfnamefont {Vladimir}\
  \bibnamefont {Fal'ko}},\ }\bibfield  {title} {\enquote {\bibinfo {title}
  {{Quartet states in two-electron quantum dots in bilayer graphene}},}\ }\href
  {\doibase 10.1103/PhysRevB.101.235423} {\bibfield  {journal} {\bibinfo
  {journal} {Phys. Rev. B}\ }\textbf {\bibinfo {volume} {101}},\ \bibinfo
  {pages} {235423} (\bibinfo {year} {2020})}\BibitemShut {NoStop}%
\bibitem [{\citenamefont {Konschuh}\ \emph {et~al.}(2012)\citenamefont
  {Konschuh}, \citenamefont {Gmitra}, \citenamefont {Kochan},\ and\
  \citenamefont {Fabian}}]{Konschuh2012Mar}%
  \BibitemOpen
  \bibfield  {author} {\bibinfo {author} {\bibfnamefont {S.}~\bibnamefont
  {Konschuh}}, \bibinfo {author} {\bibfnamefont {M.}~\bibnamefont {Gmitra}},
  \bibinfo {author} {\bibfnamefont {D.}~\bibnamefont {Kochan}}, \ and\ \bibinfo
  {author} {\bibfnamefont {J.}~\bibnamefont {Fabian}},\ }\bibfield  {title}
  {\enquote {\bibinfo {title} {{Theory of spin-orbit coupling in bilayer
  graphene}},}\ }\href {\doibase 10.1103/PhysRevB.85.115423} {\bibfield
  {journal} {\bibinfo  {journal} {Phys. Rev. B}\ }\textbf {\bibinfo {volume}
  {85}},\ \bibinfo {pages} {115423} (\bibinfo {year} {2012})}\BibitemShut
  {NoStop}%
\bibitem [{\citenamefont {Kane}\ and\ \citenamefont
  {Mele}(2005)}]{Kane2005Nov}%
  \BibitemOpen
  \bibfield  {author} {\bibinfo {author} {\bibfnamefont {C.~L.}\ \bibnamefont
  {Kane}}\ and\ \bibinfo {author} {\bibfnamefont {E.~J.}\ \bibnamefont
  {Mele}},\ }\bibfield  {title} {\enquote {\bibinfo {title} {{Quantum Spin Hall
  Effect in Graphene}},}\ }\href {\doibase 10.1103/PhysRevLett.95.226801}
  {\bibfield  {journal} {\bibinfo  {journal} {Phys. Rev. Lett.}\ }\textbf
  {\bibinfo {volume} {95}},\ \bibinfo {pages} {226801} (\bibinfo {year}
  {2005})}\BibitemShut {NoStop}%
\bibitem [{\citenamefont {Min}\ \emph {et~al.}(2006)\citenamefont {Min},
  \citenamefont {Hill}, \citenamefont {Sinitsyn}, \citenamefont {Sahu},
  \citenamefont {Kleinman},\ and\ \citenamefont {MacDonald}}]{Min2006Oct}%
  \BibitemOpen
  \bibfield  {author} {\bibinfo {author} {\bibfnamefont {Hongki}\ \bibnamefont
  {Min}}, \bibinfo {author} {\bibfnamefont {J.~E.}\ \bibnamefont {Hill}},
  \bibinfo {author} {\bibfnamefont {N.~A.}\ \bibnamefont {Sinitsyn}}, \bibinfo
  {author} {\bibfnamefont {B.~R.}\ \bibnamefont {Sahu}}, \bibinfo {author}
  {\bibfnamefont {Leonard}\ \bibnamefont {Kleinman}}, \ and\ \bibinfo {author}
  {\bibfnamefont {A.~H.}\ \bibnamefont {MacDonald}},\ }\bibfield  {title}
  {\enquote {\bibinfo {title} {{Intrinsic and Rashba spin-orbit interactions in
  graphene sheets}},}\ }\href {\doibase 10.1103/PhysRevB.74.165310} {\bibfield
  {journal} {\bibinfo  {journal} {Phys. Rev. B}\ }\textbf {\bibinfo {volume}
  {74}},\ \bibinfo {pages} {165310} (\bibinfo {year} {2006})}\BibitemShut
  {NoStop}%
\bibitem [{\citenamefont {Yao}\ \emph {et~al.}(2007)\citenamefont {Yao},
  \citenamefont {Ye}, \citenamefont {Qi}, \citenamefont {Zhang},\ and\
  \citenamefont {Fang}}]{Yao2007Jan}%
  \BibitemOpen
  \bibfield  {author} {\bibinfo {author} {\bibfnamefont {Yugui}\ \bibnamefont
  {Yao}}, \bibinfo {author} {\bibfnamefont {Fei}\ \bibnamefont {Ye}}, \bibinfo
  {author} {\bibfnamefont {Xiao-Liang}\ \bibnamefont {Qi}}, \bibinfo {author}
  {\bibfnamefont {Shou-Cheng}\ \bibnamefont {Zhang}}, \ and\ \bibinfo {author}
  {\bibfnamefont {Zhong}\ \bibnamefont {Fang}},\ }\bibfield  {title} {\enquote
  {\bibinfo {title} {{Spin-orbit gap of graphene: First-principles
  calculations}},}\ }\href {\doibase 10.1103/PhysRevB.75.041401} {\bibfield
  {journal} {\bibinfo  {journal} {Phys. Rev. B}\ }\textbf {\bibinfo {volume}
  {75}},\ \bibinfo {pages} {041401} (\bibinfo {year} {2007})}\BibitemShut
  {NoStop}%
\bibitem [{\citenamefont {Zollner}\ \emph {et~al.}(2019)\citenamefont
  {Zollner}, \citenamefont {Gmitra},\ and\ \citenamefont
  {Fabian}}]{Zollner2019Mar}%
  \BibitemOpen
  \bibfield  {author} {\bibinfo {author} {\bibfnamefont {Klaus}\ \bibnamefont
  {Zollner}}, \bibinfo {author} {\bibfnamefont {Martin}\ \bibnamefont
  {Gmitra}}, \ and\ \bibinfo {author} {\bibfnamefont {Jaroslav}\ \bibnamefont
  {Fabian}},\ }\bibfield  {title} {\enquote {\bibinfo {title}
  {{Heterostructures of graphene and hBN: Electronic, spin-orbit, and spin
  relaxation properties from first principles}},}\ }\href {\doibase
  10.1103/PhysRevB.99.125151} {\bibfield  {journal} {\bibinfo  {journal} {Phys.
  Rev. B}\ }\textbf {\bibinfo {volume} {99}},\ \bibinfo {pages} {125151}
  (\bibinfo {year} {2019})}\BibitemShut {NoStop}%
\bibitem [{\citenamefont {Banszerus}\ \emph
  {et~al.}(2020{\natexlab{b}})\citenamefont {Banszerus}, \citenamefont {Frohn},
  \citenamefont {Fabian}, \citenamefont {Somanchi}, \citenamefont {Epping},
  \citenamefont {M{\ifmmode\ddot{u}\else\"{u}\fi}ller}, \citenamefont
  {Neumaier}, \citenamefont {Watanabe}, \citenamefont {Taniguchi},
  \citenamefont {Libisch}, \citenamefont {Beschoten}, \citenamefont {Hassler},\
  and\ \citenamefont {Stampfer}}]{Banszerus2020May}%
  \BibitemOpen
  \bibfield  {author} {\bibinfo {author} {\bibfnamefont {L.}~\bibnamefont
  {Banszerus}}, \bibinfo {author} {\bibfnamefont {B.}~\bibnamefont {Frohn}},
  \bibinfo {author} {\bibfnamefont {T.}~\bibnamefont {Fabian}}, \bibinfo
  {author} {\bibfnamefont {S.}~\bibnamefont {Somanchi}}, \bibinfo {author}
  {\bibfnamefont {A.}~\bibnamefont {Epping}}, \bibinfo {author} {\bibfnamefont
  {M.}~\bibnamefont {M{\ifmmode\ddot{u}\else\"{u}\fi}ller}}, \bibinfo {author}
  {\bibfnamefont {D.}~\bibnamefont {Neumaier}}, \bibinfo {author}
  {\bibfnamefont {K.}~\bibnamefont {Watanabe}}, \bibinfo {author}
  {\bibfnamefont {T.}~\bibnamefont {Taniguchi}}, \bibinfo {author}
  {\bibfnamefont {F.}~\bibnamefont {Libisch}}, \bibinfo {author} {\bibfnamefont
  {B.}~\bibnamefont {Beschoten}}, \bibinfo {author} {\bibfnamefont
  {F.}~\bibnamefont {Hassler}}, \ and\ \bibinfo {author} {\bibfnamefont
  {C.}~\bibnamefont {Stampfer}},\ }\bibfield  {title} {\enquote {\bibinfo
  {title} {{Observation of the Spin-Orbit Gap in Bilayer Graphene by
  One-Dimensional Ballistic Transport}},}\ }\href {\doibase
  10.1103/PhysRevLett.124.177701} {\bibfield  {journal} {\bibinfo  {journal}
  {Phys. Rev. Lett.}\ }\textbf {\bibinfo {volume} {124}},\ \bibinfo {pages}
  {177701} (\bibinfo {year} {2020}{\natexlab{b}})}\BibitemShut {NoStop}%
\bibitem [{\citenamefont {Sichau}\ \emph {et~al.}(2019)\citenamefont {Sichau},
  \citenamefont {Prada}, \citenamefont {Anlauf}, \citenamefont {Lyon},
  \citenamefont {Bosnjak}, \citenamefont {Tiemann},\ and\ \citenamefont
  {Blick}}]{Sichau2019Feb}%
  \BibitemOpen
  \bibfield  {author} {\bibinfo {author} {\bibfnamefont {J.}~\bibnamefont
  {Sichau}}, \bibinfo {author} {\bibfnamefont {M.}~\bibnamefont {Prada}},
  \bibinfo {author} {\bibfnamefont {T.}~\bibnamefont {Anlauf}}, \bibinfo
  {author} {\bibfnamefont {T.~J.}\ \bibnamefont {Lyon}}, \bibinfo {author}
  {\bibfnamefont {B.}~\bibnamefont {Bosnjak}}, \bibinfo {author} {\bibfnamefont
  {L.}~\bibnamefont {Tiemann}}, \ and\ \bibinfo {author} {\bibfnamefont
  {R.~H.}\ \bibnamefont {Blick}},\ }\bibfield  {title} {\enquote {\bibinfo
  {title} {{Resonance Microwave Measurements of an Intrinsic Spin-Orbit
  Coupling Gap in Graphene: A Possible Indication of a Topological State}},}\
  }\href {\doibase 10.1103/PhysRevLett.122.046403} {\bibfield  {journal}
  {\bibinfo  {journal} {Phys. Rev. Lett.}\ }\textbf {\bibinfo {volume} {122}},\
  \bibinfo {pages} {046403} (\bibinfo {year} {2019})}\BibitemShut {NoStop}%
\bibitem [{\citenamefont {Klinovaja}\ \emph {et~al.}(2012)\citenamefont
  {Klinovaja}, \citenamefont {Ferreira},\ and\ \citenamefont
  {Loss}}]{Klinovaja2012Dec}%
  \BibitemOpen
  \bibfield  {author} {\bibinfo {author} {\bibfnamefont {Jelena}\ \bibnamefont
  {Klinovaja}}, \bibinfo {author} {\bibfnamefont {Gerson~J.}\ \bibnamefont
  {Ferreira}}, \ and\ \bibinfo {author} {\bibfnamefont {Daniel}\ \bibnamefont
  {Loss}},\ }\bibfield  {title} {\enquote {\bibinfo {title} {{Helical states in
  curved bilayer graphene}},}\ }\href {\doibase 10.1103/PhysRevB.86.235416}
  {\bibfield  {journal} {\bibinfo  {journal} {Phys. Rev. B}\ }\textbf {\bibinfo
  {volume} {86}},\ \bibinfo {pages} {235416} (\bibinfo {year}
  {2012})}\BibitemShut {NoStop}%
\bibitem [{\citenamefont {Kuemmeth}\ \emph {et~al.}(2008)\citenamefont
  {Kuemmeth}, \citenamefont {Ilani}, \citenamefont {Ralph},\ and\ \citenamefont
  {McEuen}}]{Kuemmeth2008Mar}%
  \BibitemOpen
  \bibfield  {author} {\bibinfo {author} {\bibfnamefont {F.}~\bibnamefont
  {Kuemmeth}}, \bibinfo {author} {\bibfnamefont {S.}~\bibnamefont {Ilani}},
  \bibinfo {author} {\bibfnamefont {D.~C.}\ \bibnamefont {Ralph}}, \ and\
  \bibinfo {author} {\bibfnamefont {P.~L.}\ \bibnamefont {McEuen}},\ }\bibfield
   {title} {\enquote {\bibinfo {title} {{Coupling of spin and orbital motion of
  electrons in carbon nanotubes}},}\ }\href {\doibase 10.1038/nature06822}
  {\bibfield  {journal} {\bibinfo  {journal} {Nature}\ }\textbf {\bibinfo
  {volume} {452}},\ \bibinfo {pages} {448--452} (\bibinfo {year}
  {2008})}\BibitemShut {NoStop}%
\bibitem [{\citenamefont {Jespersen}\ \emph {et~al.}(2011)\citenamefont
  {Jespersen}, \citenamefont {Grove-Rasmussen}, \citenamefont {Paaske},
  \citenamefont {Muraki}, \citenamefont {Fujisawa}, \citenamefont
  {Nyg{\aa}rd},\ and\ \citenamefont {Flensberg}}]{Jespersen2011Apr}%
  \BibitemOpen
  \bibfield  {author} {\bibinfo {author} {\bibfnamefont {T.~S.}\ \bibnamefont
  {Jespersen}}, \bibinfo {author} {\bibfnamefont {K.}~\bibnamefont
  {Grove-Rasmussen}}, \bibinfo {author} {\bibfnamefont {J.}~\bibnamefont
  {Paaske}}, \bibinfo {author} {\bibfnamefont {K.}~\bibnamefont {Muraki}},
  \bibinfo {author} {\bibfnamefont {T.}~\bibnamefont {Fujisawa}}, \bibinfo
  {author} {\bibfnamefont {J.}~\bibnamefont {Nyg{\aa}rd}}, \ and\ \bibinfo
  {author} {\bibfnamefont {K.}~\bibnamefont {Flensberg}},\ }\bibfield  {title}
  {\enquote {\bibinfo {title} {{Gate-dependent spin{\textendash}orbit coupling
  in multielectron carbon nanotubes}},}\ }\href {\doibase 10.1038/nphys1880}
  {\bibfield  {journal} {\bibinfo  {journal} {Nat. Phys.}\ }\textbf {\bibinfo
  {volume} {7}},\ \bibinfo {pages} {348--353} (\bibinfo {year}
  {2011})}\BibitemShut {NoStop}%
\bibitem [{\citenamefont {Banszerus}\ \emph {et~al.}(2018)\citenamefont
  {Banszerus}, \citenamefont {Frohn}, \citenamefont {Epping}, \citenamefont
  {Neumaier}, \citenamefont {Watanabe}, \citenamefont {Taniguchi},\ and\
  \citenamefont {Stampfer}}]{Banszerus2018Aug}%
  \BibitemOpen
  \bibfield  {author} {\bibinfo {author} {\bibfnamefont {L.}~\bibnamefont
  {Banszerus}}, \bibinfo {author} {\bibfnamefont {B.}~\bibnamefont {Frohn}},
  \bibinfo {author} {\bibfnamefont {A.}~\bibnamefont {Epping}}, \bibinfo
  {author} {\bibfnamefont {D.}~\bibnamefont {Neumaier}}, \bibinfo {author}
  {\bibfnamefont {K.}~\bibnamefont {Watanabe}}, \bibinfo {author}
  {\bibfnamefont {T.}~\bibnamefont {Taniguchi}}, \ and\ \bibinfo {author}
  {\bibfnamefont {C.}~\bibnamefont {Stampfer}},\ }\bibfield  {title} {\enquote
  {\bibinfo {title} {{Gate-Defined Electron{\textendash}Hole Double Dots in
  Bilayer Graphene}},}\ }\href {\doibase 10.1021/acs.nanolett.8b01303}
  {\bibfield  {journal} {\bibinfo  {journal} {Nano Lett.}\ }\textbf {\bibinfo
  {volume} {18}},\ \bibinfo {pages} {4785--4790} (\bibinfo {year}
  {2018})}\BibitemShut {NoStop}%
\bibitem [{\citenamefont {Tong}\ \emph {et~al.}(2021)\citenamefont {Tong},
  \citenamefont {Garreis}, \citenamefont {Knothe}, \citenamefont {Eich},
  \citenamefont {Sacchi}, \citenamefont {Watanabe}, \citenamefont {Taniguchi},
  \citenamefont {Fal{'}ko}, \citenamefont {Ihn}, \citenamefont {Ensslin},\ and\
  \citenamefont {Kurzmann}}]{Tong2021Jan}%
  \BibitemOpen
  \bibfield  {author} {\bibinfo {author} {\bibfnamefont {Chuyao}\ \bibnamefont
  {Tong}}, \bibinfo {author} {\bibfnamefont {Rebekka}\ \bibnamefont {Garreis}},
  \bibinfo {author} {\bibfnamefont {Angelika}\ \bibnamefont {Knothe}}, \bibinfo
  {author} {\bibfnamefont {Marius}\ \bibnamefont {Eich}}, \bibinfo {author}
  {\bibfnamefont {Agnese}\ \bibnamefont {Sacchi}}, \bibinfo {author}
  {\bibfnamefont {Kenji}\ \bibnamefont {Watanabe}}, \bibinfo {author}
  {\bibfnamefont {Takashi}\ \bibnamefont {Taniguchi}}, \bibinfo {author}
  {\bibfnamefont {Vladimir}\ \bibnamefont {Fal{'}ko}}, \bibinfo {author}
  {\bibfnamefont {Thomas}\ \bibnamefont {Ihn}}, \bibinfo {author}
  {\bibfnamefont {Klaus}\ \bibnamefont {Ensslin}}, \ and\ \bibinfo {author}
  {\bibfnamefont {Annika}\ \bibnamefont {Kurzmann}},\ }\bibfield  {title}
  {\enquote {\bibinfo {title} {{Tunable Valley Splitting and Bipolar Operation
  in Graphene Quantum Dots}},}\ }\href {\doibase 10.1021/acs.nanolett.0c04343}
  {\bibfield  {journal} {\bibinfo  {journal} {Nano Lett.}\ }\textbf {\bibinfo
  {volume} {21}},\ \bibinfo {pages} {1068--1073} (\bibinfo {year}
  {2021})}\BibitemShut {NoStop}%
\bibitem [{\citenamefont {Zhang}\ \emph {et~al.}(2009)\citenamefont {Zhang},
  \citenamefont {Tang}, \citenamefont {Girit}, \citenamefont {Hao},
  \citenamefont {Martin}, \citenamefont {Zettl}, \citenamefont {Crommie},
  \citenamefont {Shen},\ and\ \citenamefont {Wang}}]{Zhang2009Jun}%
  \BibitemOpen
  \bibfield  {author} {\bibinfo {author} {\bibfnamefont {Yuanbo}\ \bibnamefont
  {Zhang}}, \bibinfo {author} {\bibfnamefont {Tsung-Ta}\ \bibnamefont {Tang}},
  \bibinfo {author} {\bibfnamefont {Caglar}\ \bibnamefont {Girit}}, \bibinfo
  {author} {\bibfnamefont {Zhao}\ \bibnamefont {Hao}}, \bibinfo {author}
  {\bibfnamefont {Michael~C.}\ \bibnamefont {Martin}}, \bibinfo {author}
  {\bibfnamefont {Alex}\ \bibnamefont {Zettl}}, \bibinfo {author}
  {\bibfnamefont {Michael~F.}\ \bibnamefont {Crommie}}, \bibinfo {author}
  {\bibfnamefont {Y.~Ron}\ \bibnamefont {Shen}}, \ and\ \bibinfo {author}
  {\bibfnamefont {Feng}\ \bibnamefont {Wang}},\ }\bibfield  {title} {\enquote
  {\bibinfo {title} {{Direct observation of a widely tunable bandgap in bilayer
  graphene}},}\ }\href {\doibase 10.1038/nature08105} {\bibfield  {journal}
  {\bibinfo  {journal} {Nature}\ }\textbf {\bibinfo {volume} {459}},\ \bibinfo
  {pages} {820--823} (\bibinfo {year} {2009})}\BibitemShut {NoStop}%
\bibitem [{\citenamefont {Oostinga}\ \emph {et~al.}(2007)\citenamefont
  {Oostinga}, \citenamefont {Heersche}, \citenamefont {Liu}, \citenamefont
  {Morpurgo},\ and\ \citenamefont {Vandersypen}}]{Oos2007Dec}%
  \BibitemOpen
  \bibfield  {author} {\bibinfo {author} {\bibfnamefont {Jeroen~B.}\
  \bibnamefont {Oostinga}}, \bibinfo {author} {\bibfnamefont {Hubert~B.}\
  \bibnamefont {Heersche}}, \bibinfo {author} {\bibfnamefont {Xinglan}\
  \bibnamefont {Liu}}, \bibinfo {author} {\bibfnamefont {Alberto~F.}\
  \bibnamefont {Morpurgo}}, \ and\ \bibinfo {author} {\bibfnamefont {Lieven
  M.~K.}\ \bibnamefont {Vandersypen}},\ }\bibfield  {title} {\enquote {\bibinfo
  {title} {{Gate-induced insulating state in bilayer graphene devices}},}\
  }\href {\doibase 10.1038/nmat2082} {\bibfield  {journal} {\bibinfo  {journal}
  {Nat. Mater.}\ }\textbf {\bibinfo {volume} {7}},\ \bibinfo {pages} {151--157}
  (\bibinfo {year} {2007})}\BibitemShut {NoStop}%
\bibitem [{Note1()}]{Note1}%
  \BibitemOpen
  \bibinfo {note} {In the single electron regime, the QD transitions (chemical
  potentials) are equivalent to the single particle energies.}\BibitemShut
  {Stop}%
\bibitem [{Note2()}]{Note2}%
  \BibitemOpen
  \bibinfo {note} {We consider the combined tunneling rate, $\Gamma _\protect
  \mathrm {comb}$ to be limited by the interdot tunneling rate, $\Gamma
  _\protect \mathrm {m}$: $I/e=\Gamma _\protect \mathrm {comb}\approx \Gamma
  _\protect \mathrm {m}$, \protect \leavevmode {\protect \color {black}{}where
  $I$ is the current through the DQD device and $e$ the elementary
  charge.}}\BibitemShut {Stop}%
\bibitem [{\citenamefont {Reichardt}\ and\ \citenamefont
  {Stampfer}(2016)}]{Reichardt2016Jun}%
  \BibitemOpen
  \bibfield  {author} {\bibinfo {author} {\bibfnamefont {Sven}\ \bibnamefont
  {Reichardt}}\ and\ \bibinfo {author} {\bibfnamefont {Christoph}\ \bibnamefont
  {Stampfer}},\ }\bibfield  {title} {\enquote {\bibinfo {title} {{Modeling
  charge relaxation in graphene quantum dots induced by electron-phonon
  interaction}},}\ }\href {\doibase 10.1103/PhysRevB.93.245423} {\bibfield
  {journal} {\bibinfo  {journal} {Phys. Rev. B}\ }\textbf {\bibinfo {volume}
  {93}},\ \bibinfo {pages} {245423} (\bibinfo {year} {2016})}\BibitemShut
  {NoStop}%
\bibitem [{\citenamefont {Lyon}\ \emph {et~al.}(2017)\citenamefont {Lyon},
  \citenamefont {Sichau}, \citenamefont {Dorn}, \citenamefont {Centeno},
  \citenamefont {Pesquera}, \citenamefont {Zurutuza},\ and\ \citenamefont
  {Blick}}]{Lyon2017Aug}%
  \BibitemOpen
  \bibfield  {author} {\bibinfo {author} {\bibfnamefont {T.~J.}\ \bibnamefont
  {Lyon}}, \bibinfo {author} {\bibfnamefont {J.}~\bibnamefont {Sichau}},
  \bibinfo {author} {\bibfnamefont {A.}~\bibnamefont {Dorn}}, \bibinfo {author}
  {\bibfnamefont {A.}~\bibnamefont {Centeno}}, \bibinfo {author} {\bibfnamefont
  {A.}~\bibnamefont {Pesquera}}, \bibinfo {author} {\bibfnamefont
  {A.}~\bibnamefont {Zurutuza}}, \ and\ \bibinfo {author} {\bibfnamefont
  {R.~H.}\ \bibnamefont {Blick}},\ }\bibfield  {title} {\enquote {\bibinfo
  {title} {{Probing Electron Spin Resonance in Monolayer Graphene}},}\ }\href
  {\doibase 10.1103/PhysRevLett.119.066802} {\bibfield  {journal} {\bibinfo
  {journal} {Phys. Rev. Lett.}\ }\textbf {\bibinfo {volume} {119}},\ \bibinfo
  {pages} {066802} (\bibinfo {year} {2017})}\BibitemShut {NoStop}%
\bibitem [{\citenamefont {Prada}(2021)}]{Prada2021Mar}%
  \BibitemOpen
  \bibfield  {author} {\bibinfo {author} {\bibfnamefont {M.}~\bibnamefont
  {Prada}},\ }\bibfield  {title} {\enquote {\bibinfo {title} {{Angular momentum
  anisotropy of Dirac carriers: A new twist in graphene}},}\ }\href {\doibase
  10.1103/PhysRevB.103.115425} {\bibfield  {journal} {\bibinfo  {journal}
  {Phys. Rev. B}\ }\textbf {\bibinfo {volume} {103}},\ \bibinfo {pages}
  {115425} (\bibinfo {year} {2021})}\BibitemShut {NoStop}%
\bibitem [{\citenamefont {Gmitra}\ \emph {et~al.}(2013)\citenamefont {Gmitra},
  \citenamefont {Kochan},\ and\ \citenamefont {Fabian}}]{Gmitra2013Jun}%
  \BibitemOpen
  \bibfield  {author} {\bibinfo {author} {\bibfnamefont {Martin}\ \bibnamefont
  {Gmitra}}, \bibinfo {author} {\bibfnamefont {Denis}\ \bibnamefont {Kochan}},
  \ and\ \bibinfo {author} {\bibfnamefont {Jaroslav}\ \bibnamefont {Fabian}},\
  }\bibfield  {title} {\enquote {\bibinfo {title} {{Spin-Orbit Coupling in
  Hydrogenated Graphene}},}\ }\href {\doibase 10.1103/PhysRevLett.110.246602}
  {\bibfield  {journal} {\bibinfo  {journal} {Phys. Rev. Lett.}\ }\textbf
  {\bibinfo {volume} {110}},\ \bibinfo {pages} {246602} (\bibinfo {year}
  {2013})}\BibitemShut {NoStop}%
\bibitem [{\citenamefont {Ochoa}\ \emph {et~al.}(2012)\citenamefont {Ochoa},
  \citenamefont {Castro~Neto}, \citenamefont {Fal'ko},\ and\ \citenamefont
  {Guinea}}]{Ochoa2012Dec}%
  \BibitemOpen
  \bibfield  {author} {\bibinfo {author} {\bibfnamefont {H.}~\bibnamefont
  {Ochoa}}, \bibinfo {author} {\bibfnamefont {A.~H.}\ \bibnamefont
  {Castro~Neto}}, \bibinfo {author} {\bibfnamefont {V.~I.}\ \bibnamefont
  {Fal'ko}}, \ and\ \bibinfo {author} {\bibfnamefont {F.}~\bibnamefont
  {Guinea}},\ }\bibfield  {title} {\enquote {\bibinfo {title} {{Spin-orbit
  coupling assisted by flexural phonons in graphene}},}\ }\href {\doibase
  10.1103/PhysRevB.86.245411} {\bibfield  {journal} {\bibinfo  {journal} {Phys.
  Rev. B}\ }\textbf {\bibinfo {volume} {86}},\ \bibinfo {pages} {245411}
  (\bibinfo {year} {2012})}\BibitemShut {NoStop}%
\bibitem [{\citenamefont {Island}\ \emph {et~al.}(2019)\citenamefont {Island},
  \citenamefont {Cui}, \citenamefont {Lewandowski}, \citenamefont {Khoo},
  \citenamefont {Spanton}, \citenamefont {Zhou}, \citenamefont {Rhodes},
  \citenamefont {Hone}, \citenamefont {Taniguchi}, \citenamefont {Watanabe},
  \citenamefont {Levitov}, \citenamefont {Zaletel},\ and\ \citenamefont
  {Young}}]{Island2019Jul}%
  \BibitemOpen
  \bibfield  {author} {\bibinfo {author} {\bibfnamefont {J.~O.}\ \bibnamefont
  {Island}}, \bibinfo {author} {\bibfnamefont {X.}~\bibnamefont {Cui}},
  \bibinfo {author} {\bibfnamefont {C.}~\bibnamefont {Lewandowski}}, \bibinfo
  {author} {\bibfnamefont {J.~Y.}\ \bibnamefont {Khoo}}, \bibinfo {author}
  {\bibfnamefont {E.~M.}\ \bibnamefont {Spanton}}, \bibinfo {author}
  {\bibfnamefont {H.}~\bibnamefont {Zhou}}, \bibinfo {author} {\bibfnamefont
  {D.}~\bibnamefont {Rhodes}}, \bibinfo {author} {\bibfnamefont {J.~C.}\
  \bibnamefont {Hone}}, \bibinfo {author} {\bibfnamefont {T.}~\bibnamefont
  {Taniguchi}}, \bibinfo {author} {\bibfnamefont {K.}~\bibnamefont {Watanabe}},
  \bibinfo {author} {\bibfnamefont {L.~S.}\ \bibnamefont {Levitov}}, \bibinfo
  {author} {\bibfnamefont {M.~P.}\ \bibnamefont {Zaletel}}, \ and\ \bibinfo
  {author} {\bibfnamefont {A.~F.}\ \bibnamefont {Young}},\ }\bibfield  {title}
  {\enquote {\bibinfo {title} {{Spin{\textendash}orbit-driven band inversion in
  bilayer graphene by the van der Waals proximity effect}},}\ }\href {\doibase
  10.1038/s41586-019-1304-2} {\bibfield  {journal} {\bibinfo  {journal}
  {Nature}\ }\textbf {\bibinfo {volume} {571}},\ \bibinfo {pages} {85--89}
  (\bibinfo {year} {2019})}\BibitemShut {NoStop}%
\bibitem [{\citenamefont {Albrecht}\ \emph {et~al.}(2017)\citenamefont
  {Albrecht}, \citenamefont {Moers},\ and\ \citenamefont
  {Hermanns}}]{Albrecht2017May}%
  \BibitemOpen
  \bibfield  {author} {\bibinfo {author} {\bibfnamefont {Wolfgang}\
  \bibnamefont {Albrecht}}, \bibinfo {author} {\bibfnamefont {Juergen}\
  \bibnamefont {Moers}}, \ and\ \bibinfo {author} {\bibfnamefont {Bernd}\
  \bibnamefont {Hermanns}},\ }\bibfield  {title} {\enquote {\bibinfo {title}
  {{HNF - Helmholtz Nano Facility}},}\ }\href {\doibase 10.17815/jlsrf-3-158}
  {\bibfield  {journal} {\bibinfo  {journal} {Journal of Large-Scale Research
  Facilities}\ }\textbf {\bibinfo {volume} {3}},\ \bibinfo {pages} {112}
  (\bibinfo {year} {2017})}\BibitemShut {NoStop}%
\end{thebibliography}%
	
	\newpage
	
	\begin{figure}[!thb]
		\centering
		\includegraphics[draft=false,keepaspectratio=true,clip,width=0.5\linewidth]{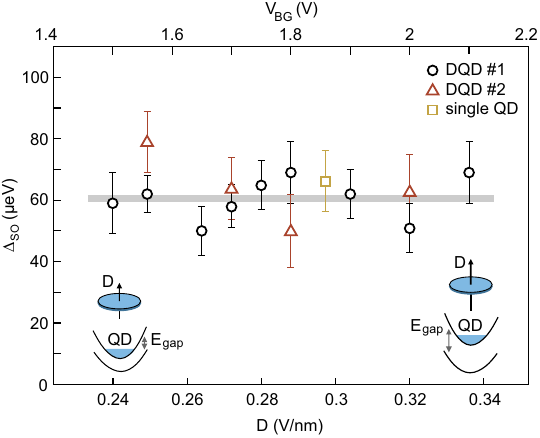}
		\new{\caption[Fig05]{ Spin-orbit gap $\Delta_\mathrm{SO}$ as function of applied electric displacement field, $D$. The black data points correspond to the device presented in the main manuscript (DQD \#1) and the red data points to the second device (DQD \#2) shown in the supplementary information. The yellow data point is obtained from a single quantum dot (single QD) formed in a third device (see supplementary information).
				Over the entire displacement field range, $\Delta_\mathrm{SO} \approx$ 60~$\mu$eV remains constant within the margin of uncertainty (see gray horizontal line). Note that the
				$V_\mathrm{BG}$-scale (upper horizontal axis) is valid only for the two double QD devices but slightly off for the single QD. Inset: Schematic illustration of the band structure of BLG with a small and a large applied out-of-plane displacement field, which breaks the inversion symmetry and leads to the opening of a band gap ($E_\mathrm{gap}$) while $\Delta_\mathrm{SO}$ remains unaffected.}}
		\label{f5}
	\end{figure}

\end{document}

\author{L.~Banszerus}
\thanks{These two authors contributed equally.}
\affiliation{JARA-FIT and 2nd Institute of Physics, RWTH Aachen University, 52074 Aachen, Germany,~EU}%
\affiliation{Peter Gr\"unberg Institute  (PGI-9), Forschungszentrum J\"ulich, 52425 J\"ulich,~Germany,~EU}

\author{S.~M\"oller}
\thanks{These two authors contributed equally.}
\affiliation{JARA-FIT and 2nd Institute of Physics, RWTH Aachen University, 52074 Aachen, Germany,~EU}%
\affiliation{Peter Gr\"unberg Institute  (PGI-9), Forschungszentrum J\"ulich, 52425 J\"ulich,~Germany,~EU}
\author{C.~Steiner}
\author{E.~Icking}
\affiliation{JARA-FIT and 2nd Institute of Physics, RWTH Aachen University, 52074 Aachen, Germany,~EU}%
\affiliation{Peter Gr\"unberg Institute  (PGI-9), Forschungszentrum J\"ulich, 52425 J\"ulich,~Germany,~EU}

\author{S.~Trellenkamp}
\author{F.~Lentz}
\affiliation{Helmholtz Nano Facility, Forschungszentrum J\"ulich, 52425 J\"ulich,~Germany,~EU}

\author{K.~Watanabe}
\affiliation{Research Center for Functional Materials, 
National Institute for Materials Science, 1-1 Namiki, Tsukuba 305-0044, Japan}
\author{T.~Taniguchi}
\affiliation{International Center for Materials Nanoarchitectonics, 
National Institute for Materials Science,  1-1 Namiki, Tsukuba 305-0044, Japan}%

\author{C.~Volk}
\author{C.~Stampfer}
\email{stampfer@physik.rwth-aachen.de}
\affiliation{JARA-FIT and 2nd Institute of Physics, RWTH Aachen University, 52074 Aachen, Germany,~EU}%
\affiliation{Peter Gr\"unberg Institute  (PGI-9), Forschungszentrum J\"ulich, 52425 J\"ulich,~Germany,~EU}%

\title{Supporting information: Spin-valley coupling in single-electron bilayer graphene quantum dots}

\date{\today}% It is always \today, today,
\maketitle

\section{Details on sample design and tuning towards a single electron DQD} 

 \begin{figure}[!thb]
\includegraphics[draft=false,keepaspectratio=true,clip,width=0.8\linewidth]{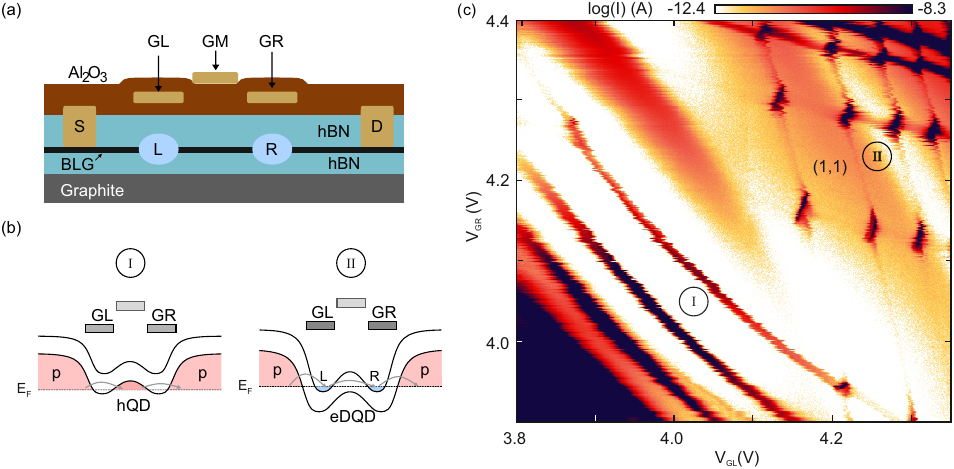}
\caption[Fig01]{
\textbf{(a)} Schematic cross section through the device. BLG is encapsulated between two flakes of hBN and placed on a graphite gate. Two Cr/Au contacts (S and D) are evaporated onto the BLG. Two layers of Cr/Au FGs are used to form a DQD. Atomic layer deposited Al$_2$O$_3$ serves as a gate dielectric. \textbf{(b)} Band alignment of the single QD and DQD regimes. \textbf{(c)} Charge stability diagram of GL versus GR showing the that the triple point considered in the main text is indeed the (0,1)-(1,0) charge transition. }
\label{f1}
\end{figure}
\newpage

\section{Measurements on a second double quantum dot} 

 \begin{figure}[!thb]
\includegraphics[draft=false,keepaspectratio=true,clip,width=\linewidth]{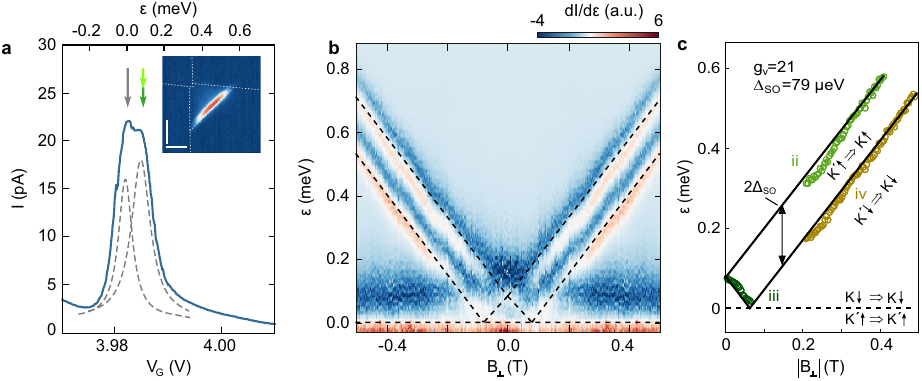}
\caption[Fig01]{ Measurements on a second DQD formed using a different set of gate fingers. \textbf{a} Detuning line cut at zero magnetic field (similar measurement as in Fig.~3d of the main text). Two resonances, one corresponding to transition (i) and one corresponding to transitions (ii) and (iii) are observed. The inset shows the the (0,1)-(1,0) triple point, the scale bars correspond to a difference in gate voltage of 20~mV. \textbf{b} Magnetotransport measurements showing $dI/d\varepsilon$ as function of detuning energy and out-of-plane $B$-field. The dashed lines mark the observed interdot transitions. \textbf{c} Extracted transition energies as function of $B_\perp$ highlighting the transitions (ii), (iii) and (iv) (see labels).}
\label{f3}
\end{figure}
\newpage

\section{Estimation of  $\Delta_\mathrm{KK'}$} 

 \begin{figure}[!thb]
\includegraphics[draft=false,keepaspectratio=true,clip,width=0.7\linewidth]{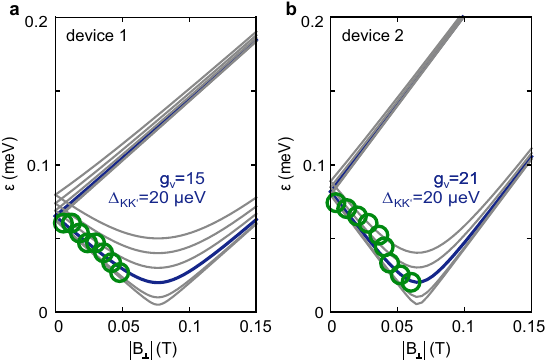}
\caption[Fig01]{ Estimation of the intervalley coupling, $\Delta_\mathrm{KK'}$ for the first (\textbf{a}) and second (\textbf{b}) double quantum dot measured. The blue lines mark the transition energies, where $\Delta_\mathrm{KK'}=20~\mu$eV is assumed, which marks an upper bound for $\Delta_\mathrm{KK'}$ in both devices. 
}
\label{f3}
\end{figure}

\newpage

\section{Reference Measurements perfomed on single QDs}

\begin{figure}[!thb]
\includegraphics[draft=false,keepaspectratio=true,clip,width=0.7\linewidth]{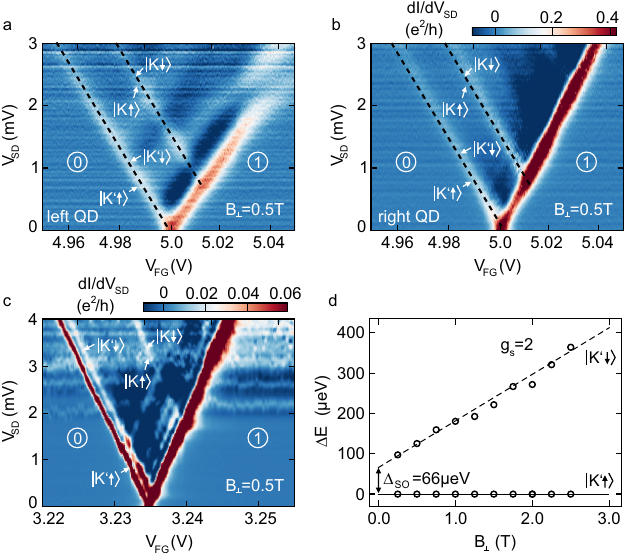}
\caption[Fig04]{\textbf{a} Finite bias spectroscopy measurement of the single electron spectrum of the left (\textbf{a}) and right (\textbf{b}) single QD of the device discussed in the main text, recorded at a perpendicular magnetic field of 0.5~T. The valley splitting due to the perpendicular field can be resolved in the two single QDs and very similar for both dots. The tunnel broadening of the Coulomb resonances do not allow to resolve the spin splitting. \textbf{c} Finite bias spectroscopy measurement performed on a different device on another chip with a design of more opaque tunneling barriers/less tunnel broadening, recorded at a finite perpendicular magnetic field of $B_\perp=$0.5~T. Here, the Zeeman spin splitting can be observed. \textbf{d} From similar measurement as depicted in c, recorded for various magnetic fields we extract the Zeeman splitting between $\ket{K'\uparrow}$ and $\ket{K'\downarrow}$. From the data, we extract a spin g-factor of $g_s=2$ and $\Delta_\mathrm{SO}=66\pm8~\mu$eV.}
\label{f3}
\end{figure}

\bibliography{Literature}